\documentclass[fleqn,usenatbib]{mnras}

\usepackage{newtxtext,newtxmath}

\usepackage[T1]{fontenc}
\usepackage{ae,aecompl}
\usepackage[utf8]{inputenc}

\usepackage{graphicx}	

\usepackage{amsmath}	
\usepackage{amssymb}	
\usepackage{comment}
\usepackage{siunitx}
\usepackage{booktabs}
\usepackage{gensymb}
\usepackage{threeparttable} 

\usepackage{xcolor}

\renewcommand{\d}{\mathrm{d}}

\let\oldnabla\nabla
\renewcommand{\nabla}{\vec{\oldnabla}}

\def\be{\begin{equation}}
\def\ee{\end{equation}}
\newcommand\code[1]{\textsc{\MakeLowercase{#1}}}

\def\gsim{\lower.5ex\hbox{\gtsima}} 
\def\lsim{\lower.5ex\hbox{\ltsima}} 
\def\gtsima{$\; \buildrel > \over \sim \;$} 
\def\ltsima{$\; \buildrel < \over \sim \;$} \def\gsim{\lower.5ex\hbox{\gtsima}} 
\def\lsim{\lower.5ex\hbox{\ltsima}} 
\def\simgt{\lower.5ex\hbox{\gtsima}} 
\def\simlt{\lower.5ex\hbox{\ltsima}}

\def\msun{{\rm M}_{\odot}}
\def\lsun{{\rm L}_{\odot}}

\def\kms{\,\rm km\,s^{-1}}

\def\ergs{{\rm erg}\,{\rm s}^{-1}}

\def\S*{$\Sigma_{\rm SFR}$}

\def\cMpc{{\rm cMpc}}
\def\cGpc{{\rm cGpc}}
\def\cMpch{{\rm cMpc} \,h^{-1}} 

\usepackage{newtxtext,newtxmath}

\usepackage[T1]{fontenc}

\DeclareRobustCommand{\VAN}[3]{#2}
\let\VANthebibliography\thebibliography
\def\thebibliography{\DeclareRobustCommand{\VAN}[3]{##3}\VANthebibliography}

\usepackage{graphicx}	
\usepackage{amsmath}	

\defcitealias{shen2007}{S07}
\def\citeS07{\citetalias{shen2007}}

\defcitealias{eftekharzadeh2015}{E15}
\def\citeE15{\citetalias{eftekharzadeh2015}}

\defcitealias{kulkarni2019}{K19}
\def\citeK19{\citetalias{kulkarni2019}}

\title[Interpreting quasar clustering at $z\approx4$]{Revisiting the Extreme Clustering of $z \approx 4$ Quasars with Large Volume Cosmological Simulations}

\author[Pizzati et al.]{Elia Pizzati$^{1}$\thanks{\href{mailto:pizzati@strw.leidenuniv.nl}{pizzati@strw.leidenuniv.nl}},
Joseph F. Hennawi$^{1,2}$,
Joop Schaye$^{1}$,
Matthieu Schaller$^{3,1}$
\\
$^{1}$ Leiden Observatory, Leiden University, P.O. Box 9513, 2300 RA Leiden,
The Netherlands\\
$^{2}$ Department of Physics, University of California, Santa Barbara, CA 93106, USA\\
$^{3}$ Lorentz Institute for Theoretical Physics, Leiden University, PO Box 9506, NL-2300 RA Leiden, The Netherlands\\
}

\date{Accepted XXX. Received YYY; in original form ZZZ}

\pubyear{2023}

\begin{document}
\label{firstpage}
\pagerange{\pageref{firstpage}--\pageref{lastpage}}
\maketitle

\begin{abstract}
Observations from wide-field quasar surveys indicate that the quasar auto-correlation length increases dramatically from $z\approx2.5$ to $z\approx4$. This large clustering amplitude at $z\approx4$ has proven hard to interpret theoretically, as it implies that quasars are hosted by the most massive dark matter halos residing in 
the most extreme environments at that redshift. In this work, we present a model that simultaneously reproduces both the observed quasar auto-correlation and quasar luminosity functions. The spatial distribution of halos and their relative abundance are obtained via a novel method that computes the halo mass and halo cross-correlation functions by combining multiple large-volume dark-matter-only cosmological simulations with different box sizes and resolutions.  Armed with these halo properties, our model exploits the conditional luminosity function framework to
describe the stochastic relationship between quasar luminosity, $L$,  and halo mass, $M$.  Assuming a simple power-law relation $L\propto M^\gamma$ with log-normal scatter, $\sigma$, we are able to reproduce observations at $z\sim 4$ and find that: (a) the quasar luminosity-halo mass relation is highly non-linear ($\gamma\gtrsim2$), with very little scatter ($\sigma \lesssim 0.3$ dex); (b) luminous quasars ($\log_{10} L/\ergs \gtrsim 46.5-47$) are hosted by halos with mass $\log_{10} M/\msun \gtrsim 13-13.5$; and (c) the implied duty cycle for quasar activity approaches unity  ($\varepsilon_{\rm DC}\approx10-60\%$). We also consider observations at $z\approx2.5$ and find that the quasar luminosity-halo mass relation evolves significantly with cosmic time, implying a rapid change in quasar host halo masses and duty cycles, which in turn suggests concurrent evolution in black hole scaling relations and/or accretion efficiency. 
\end{abstract}




\section{Introduction}\label{sec:introduction}

Quasars are extreme manifestations of the 
supermassive black holes (SMBHs) that are thought to reside at the center of almost every massive galaxy \citep[e.g.,][]{salpeter1964, zeldovich_1964, lynden_bell1969, magorrian_1998, ferrarese_merritt2000, kormendy_ho2013}. Investigating the characteristics of these luminous objects has been an active area of research for more than half a century \citep[][]{schmidt_1963}{}{}. 
In the last few years, it has become possible to trace their evolution up to redshift $z\approx7$ (\citealt{Yang_2020a, Banados_2018, Wang_2021}; see also \citealt{fan2022} for a review). 
Understanding the properties of quasars such as their abundance, luminosity, and spatial distribution, as well as their evolution with redshift, is a key step in order to study the interplay between supermassive black holes, their host galaxies, and the intergalactic medium (IGM) over cosmic time.

In particular, measuring the clustering of quasars is crucial for gaining information on the large-scale environment in which these objects reside \citep[][]{efstathiou_rees1998,cole_kaiser1989}. Like their host halos, quasars are biased tracers of the underlying distribution of dark matter \citep[e.g.,][]{kaiser1984, bardeen1986}. 
For this reason, obtaining an estimate for the linear bias factor of quasars (e.g., by measuring the quasar auto-correlation function) makes it possible to infer the characteristic masses of the halos hosting active quasars. In turn, these masses can shed light on the large-scale environment that quasars inhabit, and -- by comparing the number density of quasars with that of the hosting halos -- on the fraction of time SMBHs are shining as active quasars \citep[known as the quasar duty cycle; see e.g.][]{martini2001, haiman_hui2001, martini_2004}.

Thanks to large-sky surveys such as the Sloan Digital Sky Survey \citep[SDSS,][]{york2000} and the 2dF QSO redshift survey \citep[2QZ,][]{croom2004}, measurements of large-scale quasar clustering up to $z\approx4$ have been available for more than a decade. However, a satisfactory theoretical interpretation of these data at all redshifts is still lacking. This is mainly due to the surprising evidence that the bias factor of quasars is a steep function of redshift \citep[][]{porciani_2004,croom2005,porciani_norberg2006,shen2007,ross2009,white2012,eftekharzadeh2015,mcgreer2016,yue2021,arita2023}{}{}. While in the local universe quasars trace halos in a way that is similar to optically selected galaxies, with a bias factor close to unity \citep[][]{croom2005, ross2009}, at $z\approx4$ they are the most highly clustered objects known at that epoch, with a bias factor as high as $b\approx15$ (or, equivalently, a quasar auto-correlation length of $r_0\approx24\,\cMpch$; \citealt{shen2007}, hereafter \citeS07). Such a large correlation length implies that quasars are rare objects, arising only in the most massive halos and shining for a large fraction of the Hubble time.

Several theoretical studies have tried to reproduce the results of \citeS07 at $z\approx4$. \citet{white_2008} developed a simple model for quasar demographics that builds on a linear relation between quasar luminosity and host halo mass. They showed that to match the bias measured in \citeS07, the scatter in this relation must be very small ($\lesssim0.3$ dex). This conclusion poses two fundamental problems. Firstly, such a low scatter in the quasar luminosity-halo mass relation would be very surprising.
In fact, the conventional wisdom on the coevolution of quasars and host galaxies/halos implies that there are multiple sources of scatter contributing to determining the luminosity of a quasar at fixed halo mass (the scatter in the relations between black hole mass and quasar luminosity, black hole mass and bulge mass and between bulge mass and halo mass). 
A second concern is that low scatter in the luminosity of quasars seems to be in contrast with measurements of the relative abundance of quasars at different luminosities (the so-called quasar luminosity function, QLF). It has long been established that the bright end of the QLF is well-fitted by a power-law \citep[e.g.,][]{boyle2000,richards2006}, which stands in contrast with the exponentially-declining halo mass and galaxy luminosity functions \citep[][]{press_schechter1974,schechter1976}. The easiest way to connect these different shapes is via significant scatter in the luminosity of quasars at fixed halo/galaxy mass. 
Indeed, a number of demographic models have been developed to interpret the abundance of bright quasars and link them to their host halos \citep[e.g.,][]{croton2009,conroy_white2013,fanidakis2013,veale2014,ren_trenti2020,ren_trenti2021, trinity}. All of these studies (sometimes only implicitly) explain the relatively large number of very luminous quasars by demanding a broad range of possible quasar luminosities at a given host mass so that the more abundant population of lower-mass halos can also host a significant (or even dominant; e.g., \citealt{trinityIII}) fraction of the very bright quasars. As pointed out by some of these same studies, however, the masses of the quasar hosts implied by this picture are in plain contrast with the high masses necessary to account for the \citeS07 bias measurement.

In summary, the very strong clustering measured by \citeS07 implies a very small scatter in the luminosity of quasars at a given halo mass, and this is in tension with the large scatter required by physical models of the quasar luminosity function.
A first attempt at solving the tension
was made by \citet{shankar_2010}, using a model that connects quasar luminosities and black hole masses while accounting for the growth of black holes during cosmic time in a self-consistent way.  The authors of this study tried to match simultaneously the value of the bias inferred by \citeS07 and several measurements of the QLF at $z=3-6$ \citep[][]{shankar_mathur2007, shankar2009_review}. Assuming a non-linear relation between halo mass and quasar luminosity, they find that a low value of the scatter in this relation can reproduce 
the measurements of the bright end of the QLF. 
Even when assuming that all massive halos contribute to the clustering of quasars (i.e., a quasar duty cycle for massive systems equal to unity), however, their prediction for the $z=4$ quasar clustering is $\approx2$ standard deviations below the value measured by \citeS07. \citet{wyithe_loeb2009} also find that the \citeS07 bias measurement cannot be reproduced when assuming that the bias of dark matter halos is solely a function of their mass, and suggest that stronger clustering could be obtained if quasar activity was sparked by recent mergers (the so-called ``assembly/merger bias'', see e.g., \citealt{furlanetto_kamionkowski2006,wetzel2009, wechsler_tinker2018}). 
However, \citet{bonoli2010} \citep[see also][]{cen2015}{}{} used the Millennium Simulation \citep[][]{millennium} to study whether recently merged massive halos were clustered more strongly than other halos of the same mass, but found no evidence for that. 

Numerous other studies have compared their predictions for the quasar clustering to the \citeS07 measurements, using a variety of different approaches such as empirical models of quasar-galaxy coevolution \citep[][]{kauffmann_haehnelt_2002,hopkins2007,croton2009, shankar2010_lowz,conroy_white2013, aversa2015, shankar2020}{}{}, semi-analytic models of galaxy formation \citep[][]{bonoli2007,fanidakis2013, oogi2016}{}{} and cosmological hydrodynamical simulations \citep[][]{degraf2012, degraf2017}{}{}. While these studies are generally successful in reproducing the quasar auto-correlation function (or, equivalently, the quasar linear bias) at lower redshift ($z\lesssim3$), none of these studies have been shown to be compatible with the strong clustering observed by \citeS07. 

In conclusion, despite the efforts that have been devoted to interpreting the auto-correlation function of quasars at high redshift, a number of questions remain open: (a) is the \citeS07 measurement compatible with the standard cosmological model in which clustering is dictated by halo mass or is something akin to assembly bias 
playing an important role? (b) What is the scatter in the quasar luminosity-halo mass relation? Can small (large) scatter be reconciled with the observed quasar luminosity function (auto-correlation function)? (c) What are the physical properties that can be inferred from 
jointly modeling the QLF and quasar clustering? Can the characteristic mass 
of host halos and the quasar duty cycle be determined precisely?

One of the reasons why we have not been able to give definitive answers to these questions in more than a decade, is that modeling the clustering of high redshift quasars is difficult. The works of \citet{white_2008}, \citet{shankar_2010}, and \citet{wyithe_loeb2009} clearly show that the results of their theoretical models are strongly dependent on the assumed functional form for the linear bias-halo mass relation. 
This is because the different analytical predictions for this relation based on linear theory  \citep[e.g.][]{mo_white1996, jing98, sheth_mo_tormen_2001} diverge significantly at masses that correspond to peaks in the density perturbations that are already very non-linear \citep[][]{barkana_loeb2001}{}{}.  
For the case considered here, 
a bias of $b\approx15$, i.e., the value measured by \citeS07 for $z\approx4$ quasars, corresponds to a value of the peak height $\nu = \delta_c/\sigma(M,z)$ -- with $\delta_c\approx 1.69$ and $\sigma^2(M,z)$ being the variance of the smoothed linear density field -- equal to $\nu\approx4-6$, depending on the specific linear bias-halo mass relation and cosmology considered. Such values are rather extreme, implying that the systems contributing to the clustering of $z\approx4$ quasars live in very rare and massive environments that depart very early from the behavior expected for a linear density field. 

Improving the accuracy of the linear bias-halo mass relation via empirical fits to cosmological N-body simulations \citep[e.g.,][]{shankar_2010, tinker2010, comparat2017} does not provide a complete solution to the problem. In fact, the key point here is that the use of the large-scale linear bias as a proxy for the clustering of quasars assumes that the measured data are on quasi-linear scales, where the distribution of quasars is related to the underlying matter distribution by a scale-independent factor. This assumption breaks down for the small scales (as low as $\approx5\,\cMpc$) and for the highly non-linear environments probed by the \citeS07 data. 
For the same reason, an approach based on the (analytical) halo model framework \citep[][]{cooray_sheth2001} would also be problematic, as the non-linear bias plays a relevant role in the transition region between the one-halo and the two-halo contributions \citep[e.g.,][]{mead_verde2021, darkquest}.

In this paper, we aim to directly reproduce the observed $z=4$ quasar auto-correlation function (\citeS07) in its entirety by making use of large-volume cosmological simulations. 
This is a challenging numerical problem: in order to model the auto-correlation function properly, we need to obtain a large statistical sample of halos with masses up to $M\approx10^{13}-10^{14}\,\msun$ (which correspond, at $z=4$, to the peak heights mentioned above, i.e., $\nu\approx4-6$). Given the fact that the mass function declines exponentially at large masses, these halos are extremely rare ($1-10\,\cGpc^{-3}$), and therefore a very large simulated volume of more than $\approx 100\,\cGpc^3$ is needed to obtain a sample of at least $\approx10^2-10^3$ massive halos, that can be used to properly model the quasar auto-correlation function even at the highest masses. This is in agreement with the fact that the comoving volume probed by the SDSS observations used by \citeS07 is around $\approx50\,\cGpc^3$. A volume larger than the observational one is necessary to build a model for the quasar auto-correlation function that has higher signal-to-noise ratio than the data. At the same time, however, we also want to resolve halos down to $M\approx10^{11}-10^{12}\,\msun$ in order to explore the different possible distributions of quasars in halos that can give rise to the observed clustering. To probe these very different halo masses, we make use of a new semi-analytical framework (Sec. \ref{sec:simulations_corr} and Appendix \ref{sec:appendix_fitting}) that allows us to employ multiple simulated boxes to widen the range of masses that can be properly modeled by our simulations. 

We employ the dark-matter-only versions of the FLAMINGO suite of cosmological simulations \citep[][]{flamingo, FlamingoII} and focus on two specific box sizes: $L=2.8\,\cGpc$ and $L=5.6\,\cGpc$. On top of reproducing the clustering measurements at $z\approx4$, we also consider the constraints coming from the relative abundance of quasars at the same redshift. In other words, we aim to match the observed quasar auto-correlation and luminosity functions simultaneously. We make use of the spatial and mass distribution of halos in the simulated volumes to build a simple quasar population model that can be directly compared with observations. In this way, we are able to investigate the predictive power of quasar observables in a $\Lambda$CDM framework and obtain physical constraints on the halo mass distribution of quasar hosts and the quasar duty cycle. We also use our model to analyze the clustering and luminosity function data at a lower redshift ($z\approx2-3$), 
where the tension between theoretical models and data is not as strong \citep[e.g.,][]{croton2009,conroy_white2013, aversa2015}. This serves as a benchmark of the validity of our model and allows us to discuss the evolution of the physical properties of quasars with redshift. 

The paper is structured as follows. In Sec. \ref{sec:methods} we discuss the basic assumptions of the model, outline the cosmological simulations employed in our work, and describe how we extract the physical quantities that are necessary to model the quasar correlation function and luminosity function simultaneously. Sec. \ref{sec:model_data} gives a brief overview of the data we compare our model with, and it provides details on the statistical methodology underlying that comparison. Sec. \ref{sec:results} presents the main results of our analysis, while Sec. \ref{sec:discussion} contains a discussion of the implications of our findings and their connections to previous work.  Conclusions are provided in Sec. \ref{sec:conclusions}.

\section{Methods} \label{sec:methods}

\begin{figure*}
	\centering
	\includegraphics[width=0.95\textwidth]{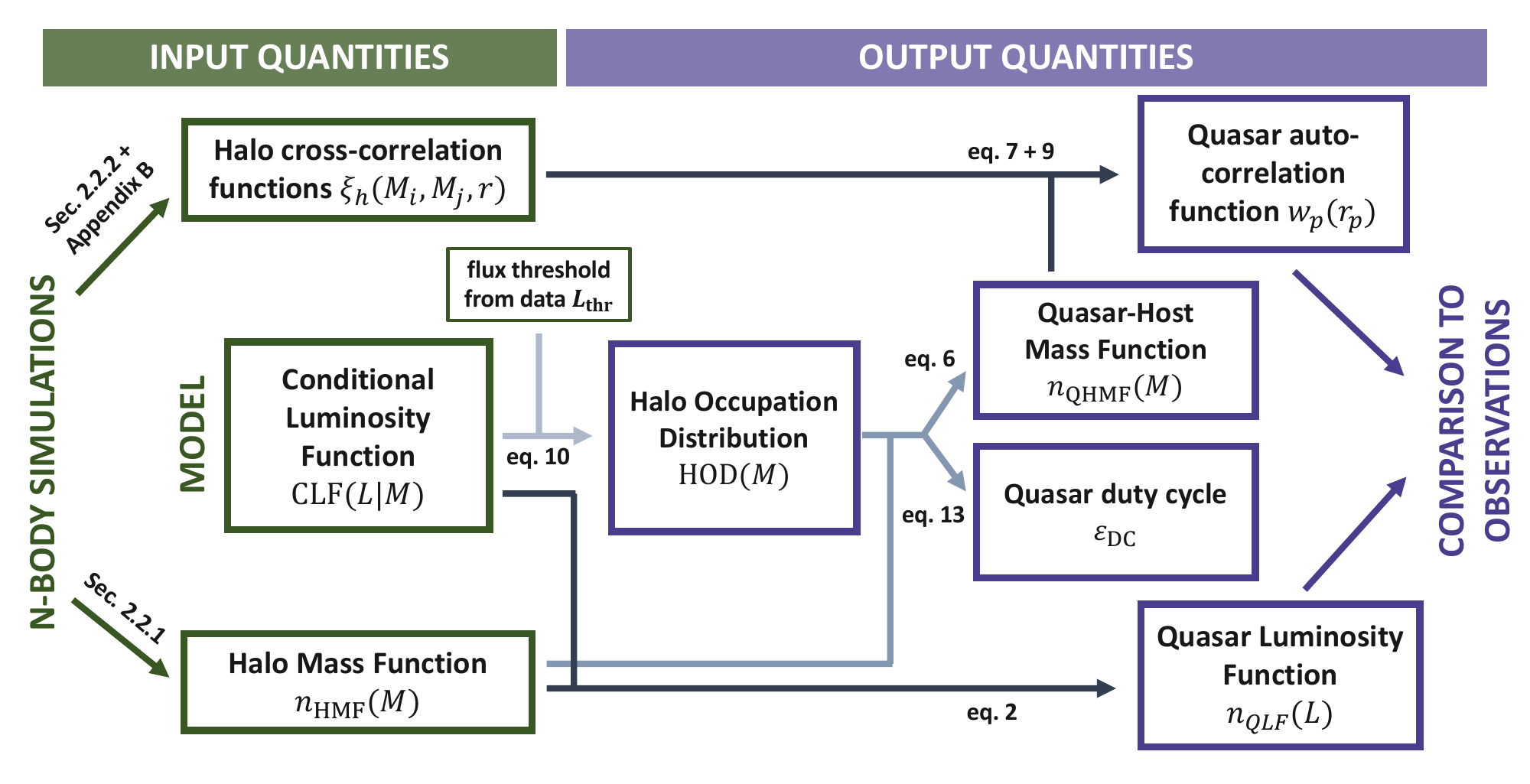}
	 \caption{ Summary of the various quantities involved in the analysis. We choose a model for the Conditional Luminosity Function (CLF) that depends on a set of free parameters. We then combine this with the halo mass function and the halo cross-correlation functions taken from the FLAMINGO cosmological simulations to obtain the two main observables of interest, the quasar luminosity function and auto-correlation function, together with other key properties such as the quasar-host mass function, the halo occupation distribution (HOD), and the quasar duty cycle. 
  \label{fig:method_summary}
 	}
\end{figure*}

In this Section, we describe our model for the distribution of quasars in space and luminosity.
We start by outlining the basic framework 
(Sec. \ref{sec:methods_basic}); then, we describe the FLAMINGO cosmological simulations and detail how we extract the mass function and the cross-correlation functions of halos (Sec. \ref{sec:methods_sim}). Figure \ref{fig:method_summary} shows a summary of the various quantities involved in our analysis, together with a reference to the equations where they are defined.

\subsection{The conditional luminosity function} \label{sec:methods_basic}

We adopt an empirical model that is agnostic to the physics underlying the quasar emission/black hole accretion mechanisms. The only assumptions we make are: (a) every halo above some mass $M_{\rm min}$ hosts a SMBH at its center, emitting at some luminosity $L$; (b) the luminosity of a SMBH depends only on the mass of the host halo, $M$. Therefore, we can employ a conditional luminosity function approach (CLF; see e.g., \citealt{yang2003, ballantyne2017a,ballantyne2017b,bhowmick2019, ren_trenti2020}) and write the 2-d distribution in the black hole 
luminosity-host halo mass plane, $n( L, M)$, as:
\begin{equation}
   n(L, M) = {\rm CLF}(L| M) \,n_\mathrm{HMF}(M),
\end{equation}
where $n_\mathrm{HMF}(M)$ is the halo mass function. 

Note that the luminosity of a SMBH, $L$, can be interpreted as either a bolometric luminosity or a luminosity in a specific band of the spectrum. The framework that we are introducing here is agnostic to this choice and can be formulated to describe the emission coming from any region of the spectrum. However, for clarity and consistency with previous work on the subject \citep[e.g.,][]{white_2008, shankar_2010, conroy_white2013, trinity}{}{}, in this paper we choose to work with bolometric luminosities. Henceforth, $L$ will always refer to the bolometric luminosity, i.e., $L\equiv L_{\rm bol}$\footnote{However, note that the data considered in this paper always refer to type I, UV-bright quasars \citep[e.g.,][]{padovani2017}{}{}. Hence, the model presented in this work describes only this specific population of active SMBHs.}.

Within this framework, the Quasar Luminosity Function 
-- $n_\mathrm{QLF}( L)$ -- is simply the marginalization of $n( L, M)$ over halo mass, $ M$:
\begin{align}
        n_\mathrm{QLF}( L) &= \int_{M_{\rm min}}^{M_{\rm max}} {\rm CLF}( L| M) \, n_\mathrm{HMF}( M)\,\d M . \label{eq:qlf}
\end{align}
Therefore, assuming that the halo mass function is known, the QLF can be easily determined once a conditional luminosity function has been adopted. The two limits of integrations, $M_{\rm min}$ and $M_{\rm max}$, represent the minimum/maximum mass of a halo that can host a SMBH. In principle, we could have SMBHs in any halos, and set this integration range to be as wide as possible. However, given that the simulations employed in our analysis span a wide but finite range of masses (Sec. \ref{sec:methods_sim}), we adopt the following limits: $\log_{10} M_{\rm min}/\msun=11.5$, and $\log_{10} M_{\rm max}/\msun=14$ ($\log_{10} M_{\rm max}/\msun=14.5$) at redshift $z=4$ ($z=2.5$). These limits enclose a range in masses that is sufficiently broad for our redshifts of interest (Sec. \ref{sec:results}), so that expanding the range would have a negligible impact on our final results. 

We use a model for the CLF in which the distribution in luminosity is log-normal at 
fixed mass \citep[see also][]{ren_trenti2020, ren_trenti2021, white2012}:
\begin{equation}
        {\rm CLF}(L| M)\, \d L= \,\frac{f_{\rm on}}{\sqrt{2\pi}\sigma}\,\exp\left(\frac{(\log_{10} L - \log_{10} L_\mathrm{c}(M))^2}{2\sigma^2}\right) \d \log_{10} L .
        \label{eq:clf_log_normal}
\end{equation}
We then assume a power-law dependence of the characteristic luminosity, $L_\mathrm{c}$, on mass:
\begin{equation}
    L_\mathrm{c}(M) = L_{\rm ref} \,\left(\frac{M}{M_{\rm ref}}\right)^\gamma,
\end{equation}
or, in terms of logarithmic quantities:
\begin{equation}
    \log_{10} L_\mathrm{c}(M) = \log_{10} L_{\rm ref} +\gamma\, \left(\log_{10} M - \log_{10} M_{\rm ref}\right),
\end{equation}
where $M_{\rm ref}$ is simply a reference mass that is associated with the reference luminosity $L_{\rm ref}$. We fix $\log_{10} M_{\rm ref}/\msun = 12.5$. 
The free parameters of the model are: $\sigma$, $L_\mathrm{ref}$, $\gamma$, and $f_\mathrm{on}$. In the following, we assume that these parameters do not depend on the other variables such as halo mass or quasar luminosity, and let them assume different values for the different redshifts we consider in Sec. \ref{sec:results}.

The factor $f_{\rm on}$ accounts for the fact that not all black holes may be active as quasars at any given time. Therefore, we are implicitly assuming that the CLF is bimodal: the first mode accounts for all luminous quasars and is log-normally distributed, whereas the second mode (not accounted for in eq. \ref{eq:clf_log_normal}) describes the behavior of the black holes that are too dim to be probed by any observations and is therefore completely irrelevant to our analysis. This bimodality in the CLF has a well-defined physical meaning: black holes are either active as luminous quasars or they are dormant, with a luminosity that is orders of magnitudes lower than any observational limits. However, it is not clear whether the luminosity distribution of black holes is indeed bimodal, or rather shows a continuum between active sources and inactive/faint ones. 
Observations of very faint quasars ($\log_{10} L/\ergs \approx 42-45$) can 
shed light on this question\footnote{Such observations become very difficult in the distant universe, as faint quasars are often outshined by their host galaxies.}. We will return to this point in Sec. \ref{sec:discussion_caveats}. 

\subsubsection{The quasar auto-correlation function}

In our framework, the correlation function of quasars is identical to the correlation function of the halos that host them, as quasars are 
temporally subsampling the underlying halo distribution. However, we have to consider that only quasars above some luminosity threshold $L_{\rm thr}$ are accounted for when measuring the correlation function in a survey. Therefore, we are effectively considering a ``biased'' halo mass distribution traced by the quasars above this luminosity threshold: we will refer to it as the ``Quasar-Host Mass Function'' (QHMF). This quantity can be expressed in terms of the halo mass function and another marginalization integral of the CLF:
\begin{equation}
    n_\mathrm{QHMF}( M|L>L_{\rm thr}) = n_\mathrm{HMF}(M)\,\int_{ L_{\rm thr}}^\infty {\rm CLF}( L| M) \, \d  L .\label{eq:qhmf}
\end{equation}
The clustering of quasars can then be determined by computing the correlation function of a sample of halos that are distributed according to $n_\mathrm{QHMF}( M|L>L_{\rm thr})$. Here, we use an approach that allows us to quickly determine the quasar auto-correlation function for different $n_\mathrm{QHMF}( M)$ distributions: we create different mass bins, and -- selecting halos in these bins -- extract the cross-correlation functions for halos with different masses from a cosmological simulation (see Sec. \ref{sec:methods_sim} for more details). Let us call these cross-correlation terms $\xi_h(M_j, M_k; r)$, with $M_{j,k}$ being the centers of the mass bins. We can then compute the quasar auto-correlation function, $\xi(r)$,  by simply weighting the cross-correlations terms, $\xi_h(M_j, M_k; r)$, according to the quasar-host mass function, $n_\mathrm{QHMF}$:
    \begin{equation}
        \xi(r) 
        = \sum_{j,k} p_j p_k \xi_h(M_j, M_k; r), \label{eq:quasar_corr_func}
    \end{equation}
where the weights $p_{j,k}$ are defined as:
\begin{equation}
     p_i = \frac{n_\mathrm{QHMF}( M_i|L>L_{\rm thr})\,\Delta M }{\int_0^{M_{\rm max}} n_\mathrm{QHMF}( M|L>L_{\rm thr})\,\d M},
\end{equation}
with $\Delta M$ being the width of the mass bins.
We present how to derive these equations in Appendix \ref{sec:cross-corr}. 

Once $\xi(r)$ is known, other related quantities such as the projected auto-correlation function, $w_\mathrm{p}(r_\mathrm{p})$, can be easily obtained by integrating along the parallel direction $\pi$. The projected auto-correlation function is relevant since it can be directly compared to observational data (see Sec. \ref{sec:obs_data})\footnote{$w_\mathrm{p}(r_\mathrm{p})$ is commonly used as a statistic for clustering measurements because it allows averaging out the contribution of redshift-space distortions to the observed 3-d correlation function. In the analysis performed here, we will always compare our model with $w_\mathrm{p}(r_\mathrm{p})$ data, and hence will not take into account the presence of redshift-space distortions. The framework presented here, however, can easily be generalized to model redshift-space correlation functions.}. Setting a maximum value for the parallel distance $\pi_{\rm max}$, which is chosen in accordance with the one used for observational data, e.g. $\pi_{\rm max}=100\,\cMpch$ for the \citeS07 measurements, the projected auto-correlation function reads:
    \begin{equation}
        w_\mathrm{p}(r_\mathrm{p}) = \int_{-\pi_{\rm max}}^{\pi_{\rm max} } \xi(r_\mathrm{p}, \pi)\,\d \pi = 2 \int_{r_\mathrm{p}}^{\sqrt{r_\mathrm{p}^2 + \pi_{\rm max}^2}} \frac{r\xi(r)}{\sqrt{r^2-r_\mathrm{p}^2}}\,\d r .\label{eq:projected_corrfunc}
    \end{equation}

\subsubsection{Halo occupation distribution and duty cycle}

From the CLF, we can extract other quantities that will be relevant to our analysis. In particular, the integral of the CLF above some threshold luminosity $L_{\rm thr}$ represents the
aggregate probability for a halo of mass $M$ to host a quasar with a luminosity above the threshold value. Therefore, it is equivalent to a Halo Occupation Distribution (HOD; see e.g., \citealt{berlind2002}):
\begin{equation}
    {\rm HOD}(M) = \frac{n_\mathrm{QHMF}( M|L>L_{\rm thr})}{n_\mathrm{HMF}(M)} = \int_{L_{\rm thr}}^\infty {\rm CLF}( L| M) \, \d  L .\label{eq:hod}
\end{equation}

The HOD is also closely related to the idea of a \textit{quasar duty cycle}. In fact, the duty cycle is defined as the fraction of active quasars (i.e., with a luminosity above the threshold) divided by the fraction of halos that are able to host these quasars. In the standard picture (e.g., \citealt{martini2001,haiman_hui2001}) this fraction is well defined, as it is implicitly assumed that there is a minimum halo mass $\Tilde{M}_{\rm min}$ above which all halos can host quasars, and only a fraction of them is active at the present moment. In other words, the QHMF is: 
\begin{equation}
    n_{\rm QHMF}(M)=\varepsilon_{\rm DC}n_{\rm HMF}(M)\Theta(\log_{10} M - \log_{10} \Tilde{M}_{\rm min}),
\end{equation}
with $\varepsilon_{\rm DC}$ being the duty cycle and $\Theta$ the Heaviside step function. 
However, this definition of the duty cycle is not well-posed in our approach. As described above, we do not assume a specific functional form for the QHMF, but rather we infer this quantity from the CLF (eq. \ref{eq:qhmf}). As a consequence, we do not define a minimum mass $\Tilde{M}_{\rm min}$ for halos to host bright quasars, but rather consider all halos and compute a probability for them to host bright quasars given their mass, $M$. This implies that, in principle, even halos with a very low mass could have a small but non-zero probability to host bright quasars. As a result, the above definition of the quasar duty cycle would return artificially small values. Therefore, we opt here for an alternative definition (see also \citealt{ren_trenti2021}): the duty cycle, $\varepsilon_{\rm DC}$, is the weighted average of the HOD above a threshold mass that is given by the median of the QHMF, $n_\mathrm{QHMF}( M|L>L_{\rm thr})$.
In other words, if we define the median of the QHMF as the mass $M_{\rm med}$ satisfying the relation:
\begin{equation}
   \int_{M_{\rm med}}^{M_{\rm max} }n_{\rm QHMF}(M) = 0.5\,\int_{M_{\rm min}}^{M_{\rm max}} n_{\rm QHMF}(M), \label{eq:qhmf_med}
\end{equation}
then $\varepsilon_{\rm DC}$ can be expressed as:
\begin{equation}
\begin{split}
        \varepsilon_{\rm DC} &= \frac{\int_{M_{\rm med}}^{M_{\rm max}}  n_\mathrm{HMF}( M)\,{\rm HOD}(M) \,  \d M}{\int_{M_{\rm med}}^{M_{\rm max}} n_\mathrm{HMF}(M)\,\d M} =\\
        &= \frac{\int_{M_{\rm med}}^{M_{\rm max}} n_\mathrm{QHMF}( M|L>L_{\rm thr})\,\d M}{\int_{M_{\rm med}}^{M_{\rm max}} n_\mathrm{HMF}( M)\,\d M}.\\
        \label{eq:duty_cycle}
        \end{split}
\end{equation}

 \begin{figure*}
	\centering
	\includegraphics[width=0.49\textwidth]{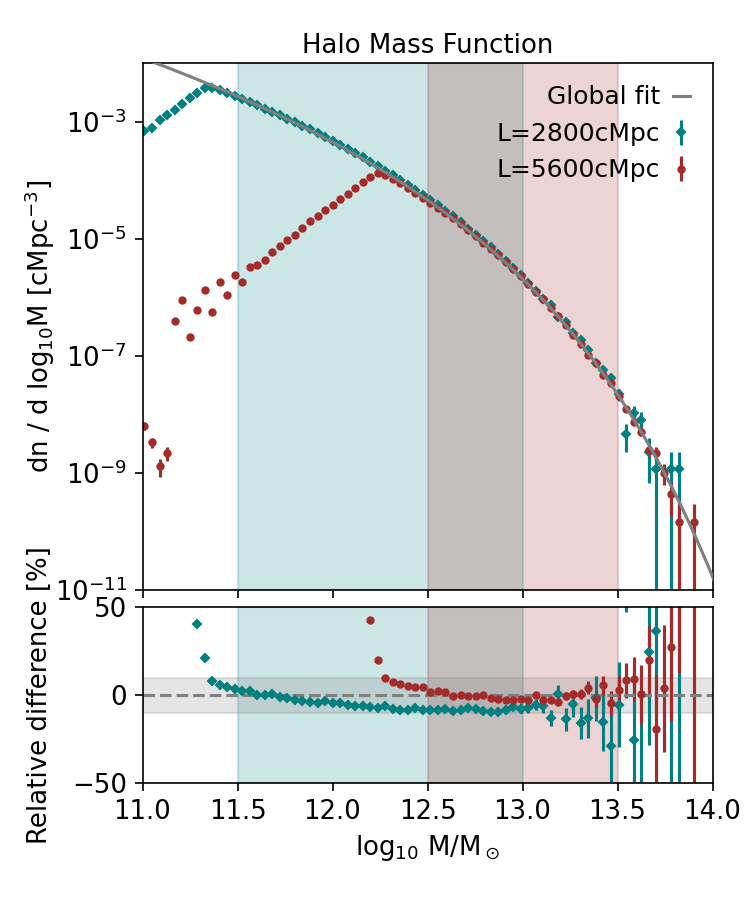}
 	\includegraphics[width=0.49\textwidth]{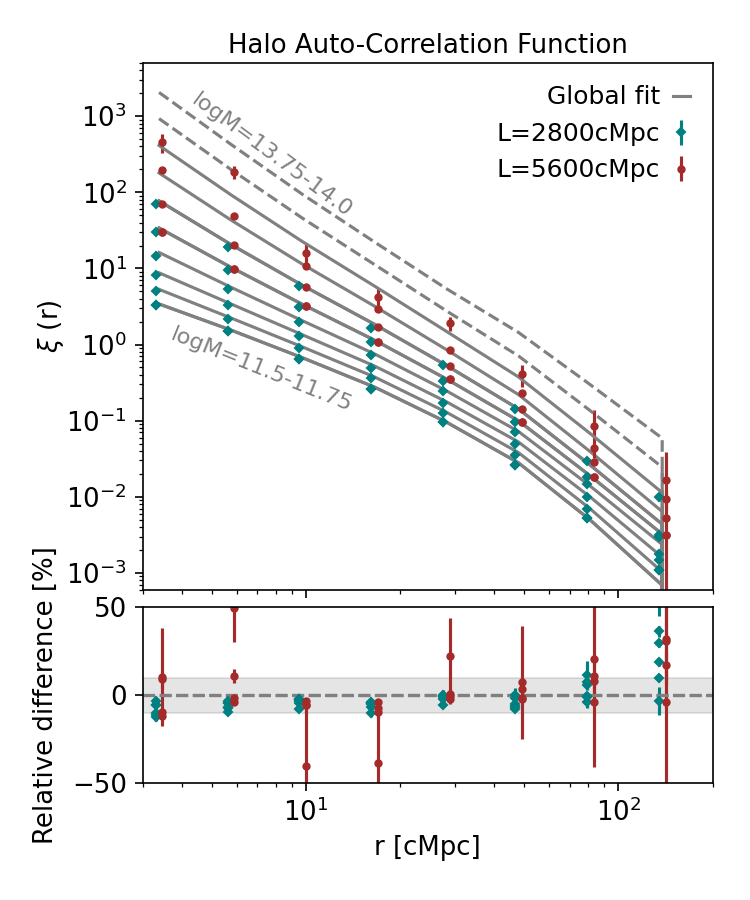}

	 \caption{{\it Left:} Halo mass function at $z=4$ from the simulations considered: $L=2800\,\cMpc$ (teal diamonds) and $L=5600\,\cMpc$ (red circles). The gray solid line represents the analytical fit to the simulations (see Sec. \ref{sec:simulations_hmf} for more details). The shaded regions highlight which masses in each simulation are considered for the fit. The bottom panel shows the relative difference between the fit and the two simulations (with the horizontal shaded grey band highlighting the $10\%$ limit). {\it Right:} Auto-correlation function of halos in different mass bins at $z=4$. We create 8 mass bins ranging from $\log_{10} M/\msun = 11.5$ to $\log_{10} M/\msun = 13.5$ and $0.25$ dex wide. Lower mass bins correspond to lower values of the correlation functions, and vice-versa. Teal diamonds refer to the $L=2800\,\cMpc$ simulation, while red circles refer to the $L=5600\,\cMpc$ one. Points are staggered in the x-direction for visualization purposes. The gray solid lines represent the fits to the auto-correlation functions (from the lowest mass bin on the bottom to the highest mass bin on top), as described in Appendix \ref{sec:appendix_fitting}. Relative differences between the fits and the simulated correlation functions are shown in the bottom panel. These differences are generally $\lesssim10\%$, with the exception of the highest mass bin considered (i.e., $\log_{10} M/\msun = 13.25-13.5$), for which the measurements are noisy due to the small number of halos in that mass range. The gray dashed lines in the top panel show extrapolations of the auto-correlation functions based on our fit for even higher mass bins ($\log_{10} M/\msun = 13.5-13.75$ and $\log_{10} M/\msun = 13.75-14.0$) where measurements from the simulations are not available. More details can be found in Sec. \ref{sec:simulations_corr} and Appendix \ref{sec:appendix_fitting}. \label{fig:simulations}
 	}
\end{figure*}

\subsection{Dark matter only simulation setup} \label{sec:methods_sim}

In the last section, we have shown how we can make use of the CLF formalism to compute the quasar luminosity and auto-correlation functions -- together with other relevant quantities such as the QHMF and the quasar duty cycle -- using two fundamental ingredients: the mass function of halos and the cross-correlation functions of halos with different masses (see Figure \ref{fig:method_summary} for a summary of this workflow). 
In this section, we provide details on how we obtain these two ingredients using the Dark-Matter-Only (DMO) version of the FLAMINGO suite of cosmological simulations.  

FLAMINGO \citep[][]{flamingo, FlamingoII} is a suite of state-of-the-art, large-volume cosmological simulations run with the N-body gravity and smooth particle hydrodynamics (SPH) solver \code{SWIFT} \citep[][]{swift}{}{}. Gravity is solved using the Fast Multiple Method \citep[][]{gravity_solver}.
The cosmology adopted in FLAMINGO is the ``3x2pt + all'' cosmology from \citet{abbott_des2022} ($\Omega_\mathrm{m} = 0.306$, $\Omega_\mathrm{b} = 0.0486$, $\sigma_8 = 0.807$, $\mathrm{H}_0 = 68.1\,\kms\,{\rm Mpc}^{-1}$, $n_\mathrm{s} = 0.967$), with a summed neutrino mass of $0.06\,\mathrm{eV}$. Massive neutrinos are included in the simulation via the $\delta f$ method of \citet{swift_neutrinos}. Initial conditions (ICs) are set using multi-fluid third-order Lagrangian
perturbation theory (3LPT). Partially fixed ICs are used to limit the impact of cosmic variance \citep[][]{angulo_pontzen2016}{}{} by setting the amplitudes of modes with
$(kL)^2<1025$ to the mean variance ($k$ is the wavenumber and
$L$ the box size).

In this work, we focus on two specific DMO simulations with box sizes $L=2800\,\cMpc$ and $L=5600\,\cMpc$, respectively. Both simulations have $5040^3$ cold dark matter (CDM) particles and $2800^3$ neutrino particles. The CDM particle masses are $M_{\rm dm}=6.72\times10^9\,\msun$ and $M_{\rm dm}=5.38\times10^{10}\,\msun$ for the $L=2800\,\cMpc$ and $L=5600\,\cMpc$ boxes, respectively. 
We focus on the DMO version of the simulations because no hydrodynamic version is available for the largest box, and because we are only interested in the spatial distribution of halos, that, in the $\Lambda {\rm CDM}$ model, is primarily dictated by gravitational interactions of dark-matter particles only. 

We identify halos in the simulated snapshots using the 6-d friends-of-friends code \code{VELOCIraptor} \citep[][]{velociraptor}. Once halos have been identified, their masses are computed using a spherical-overdensity definition based on their density profile. We perform this task using the code SOAP\footnote{\href{https://github.com/SWIFTSIM/SOAP}{https://github.com/SWIFTSIM/SOAP}}. We define the radius of a halo as the distance from the most bound particle within which the density reaches a value of 200 times the critical density of the universe ($200\rho_\mathrm{c}$). We only include central halos in the analysis and exclude the contribution of sub-halos. As discussed in Sec. \ref{sec:discussion_caveats}, we do not expect this to influence our results significantly. 

Once we have obtained a catalogue with the positions and masses of halos in the simulation at a given redshift, we can easily compute key statistical properties such as the halo mass function and the (cross-)correlation functions of halos with different masses. However, this approach is not directly suitable for our purposes. 
In fact, an important limitation of cosmological simulations is that they give reliable results only in a finite range of masses. The lower limit of this mass range is imposed by resolution: halos with fewer than $50-100$ dark-matter particles are not well resolved, and thus cannot be trusted. The upper limit, on the other hand, is set by the box size of the simulation: if the number of halos with mass greater than some threshold $M$ is small, these halos are too rare to get a reliable estimate of their statistical properties (e.g., their clustering). 

For the problem we are facing here, we need to be able to reproduce the relative abundance of halos and their spatial distribution for a vast range of masses. For this reason, employing a single halo catalogue obtained using a simulation with a fixed box size is not the optimal strategy. Instead, we use here an approach consisting of two key steps: we first compute the quantities of interest (i.e., the halo mass function and the halo cross-correlation functions) from multiple simulations with different box sizes (and mass resolutions), and then we combine these different simulations by making use of analytical fitting functions. In this way, we can predict the abundance and spatial distribution of halos for all the masses that are well captured by the different simulations considered. 

Table \ref{tab:sim} summarizes the properties of the simulations we employ. In brief, we use the two different box sizes $L=2800\,\cMpc$ and $L=5600\,\cMpc$ to study the properties of low-mass and high-mass halos, respectively. For the $2800\,\cMpc$ box, we select halos in the range of masses $\log_{10} M/\msun = 11.5-13.0$; for $L=5600\,\cMpc$, we focus on halos in the range $\log_{10} M/\msun = 12.5-13.5$. The lower limits are set to select only halos with at least $\approx50$ particles, whereas the upper limits are set to ensure overlap between the two mass ranges and to guarantee that all mass bins (up to at least $z=4$) are populated with at least $5000$ halos. In the following we describe in detail how we combine these simulations to obtain an analytical description of the halo mass function and of the cross-correlation function of halos with different masses.

\begin{table*}
\centering
\setlength{\extrarowheight}{3pt}
\caption{Summary of the different FLAMINGO cosmological simulations employed in the analysis. The ``fitting mass range'' refers to the mass range selected for the fits of the halo mass function and the cross-correlation functions (Sec. \ref{sec:simulations_hmf} and \ref{sec:simulations_corr}, respectively). The redshifts considered in the analysis are $z=4.0$ (high-redshift data; see Figure \ref{fig:simulations}), and $z=2.5$ (low-redshift data; see Figure \ref{fig:simulations_z2}).}
\begin{tabular}{ c | c c c c }
\toprule
 Box size [$\cMpc$] 
 & Number of CDM particles    & CDM particle mass [$\log_{10} \msun$]   & Fitting mass range [$\log_{10} \msun$]  & Snapshots considered ($z$)\\ 
 \midrule
2800     &    
$5040^3$     &   
$9.83$ &
$11.5-13.0$ &
$2.5,\,4.0$\\
5600    &   
$5040^3$     &   
$10.77$ &
$12.5-13.5$&
$2.5,\,4.0$\\
 \bottomrule
\end{tabular}
\label{tab:sim}
\end{table*}

\subsubsection{Fitting the halo mass function}
\label{sec:simulations_hmf}

Following \citet{tinker2008} (see also \citealt{jenkins2001,white2002, warren2006}), we write the halo mass function in terms of the peak height of the density perturbations, $\nu=\delta_c/\sigma(M,z)$, where $\delta_c\approx1.69$ is the critical linear density for collapse and $\sigma(M,z)$ is the variance of the linear density field smoothed on a scale $R(M)$ \citep[see][]{press_schechter1974,sheth_tormen1999}. According to this formalism, the mass function can be parametrized in terms of a universal function $f(\sigma)$:
\begin{equation}
    f(\sigma;A,a,b,c) = A\,\left(\left(\frac{\sigma}{b}\right)^{-a}+1\right)\,e^{-c/\sigma^2}, \label{eq:hmf_fit}
\end{equation}
where $f(\sigma)$ is related to the mass function via the expression
\begin{equation}
    \frac{\d n}{\d M}(M,z) = f(\sigma)\,\frac{\rho_{{\rm m},0}}{M}\,\frac{\d \ln \sigma^{-1}}{\d M}, \label{eq:hmf}
\end{equation}
with $\rho_{{\rm m},0}$ being the mass density at $z=0$.

We use the python package \code{colossus} \citep[][]{colossus_diemer2018} to compute the value of $\sigma(M,z)$ using the same cosmology as the FLAMINGO simulation (Sec. \ref{sec:methods_sim}). We then use $\chi^2$-minimization to find the best-fitting parameters $(A,a,b,c)$ for the analytical form of the halo mass function. We fit the number density of halos in different mass bins using halo catalogues from two different simulations, using two different (but partially overlapping) mass ranges (see Table \ref{tab:sim}). We assign Poissonian counting errors to every mass bin considered. 
We also experiment with changing these errors, and find that we achieve a better fit to the data by doubling the errors for the $L=2800\,\cMpc$ simulation, and halving the ones associated with the $L=5600\,\cMpc$ box. Note that this choice is arbitrary: our goal is not to provide a physically-motivated fit to the data, but simply to find a good analytical description of the halo mass function coming from simulations. 

Figure \ref{fig:simulations} (left panel) shows the best-fitting mass function for $z=4$, together with the data obtained from the simulations. Analogous results for $z=2.5$ are shown in Appendix \ref{sec:appendix_fitting_z2.5}. The optimal parameter values for this mass function are: $A=5.68\times10^{-5}$, $a=1.65$, $b=257$, $c=1.16$. As shown in the lower left panel of Fig. \ref{fig:simulations}, the fit provides a description of the simulated data with an accuracy of $\approx5-10\%$ up to $\log_{10} M/\msun \lesssim 13.5$. As we will discuss in Sec. \ref{sec:discussion_caveats}, this level of accuracy for the model is enough to provide a satisfactory description of the observed data. 

Finally, we note that the reason why we have performed the fitting of the halo mass functions extracted from our simulations and did not consider the best-fitting parameters provided by \citet{tinker2008} is because we found that, at $z\geq4$, differences between the \citet{tinker2008} model and our simulations were as high as $100\%$ \citep[see also][]{yung2023}{}{}.

\subsubsection{Obtaining the cross-correlation functions}
\label{sec:simulations_corr}

We want to obtain the cross-correlation functions of halos with masses $M_j$ and $M_k$, $\xi_h(M_j, M_k; r)$. In order to achieve this, we create a grid in mass and distance by considering 8 uniformly spaced bins in $\log_{10} M$, 
with a minimum halo mass of $\log_{10} M_{\rm min}/\msun=11.5$ and a maximum of $\log_{10} M_{\rm max}/\msun=13.5$, and 8 (logarithmically-spaced) bins in the radial direction with a minimum radial distance of $\log_{10} r_{\rm min}/\cMpc = 0.4$ and a maximum of $\log_{10} r_{\rm max}/\cMpc = 2.25$.  
We then use the package \code{corrfunc} \citep[][]{corrfunc} to compute the number of halo pairs in the simulated catalogues for every combination of masses and distance, together with the number of pairs obtained assuming that these halos are distributed randomly. The values of the cross-correlation terms are obtained using the \citet[][]{landy_szalay1993} estimator:
\begin{equation}
    \xi_h(M_j, M_k; r) = \xi_{j,k}(r) = \frac{D_jD_k-D_jR_k-D_kR_j+R_jR_k}{R_jR_k}, \label{eq:pair_counts}
\end{equation}
where $D_jD_k$ stands for the number of pairs of halos in the mass bin $j$ with halos in the mass bin $k$, whereas $R_jD_k$, $D_jR_k$, and $R_jR_k$ refer to the number of pairs when comparing to a random distribution of the same halos. 

We end up with 36 different cross-correlation functions -- i.e., the number of independent elements for a symmetric 64-element matrix -- which can be used to determine the quasar auto-correlation function according to eq. \ref{eq:quasar_corr_func}. 
However, once again, we must account for the fact that different simulations probe different mass ranges. We thus fit a parametric analytical function
to these cross-correlation functions in a way that allows us to combine different simulated boxes. 

Furthermore, in this case the fitting procedure has another critical purpose. Despite the large volume of the simulations employed, the number of simulated halos at the very high mass end is limited by the finite size of the box. For this reason, the obtained cross-correlation terms for the very high-mass halo pairs will suffer from significant uncertainties due to the limited sample size in the simulation. Even for the largest box we consider (i.e., $L=5600\,\cMpc$), at $z=4$ this effect starts to be significant for $\log_{10} M/\msun \approx 13.2-13.5$. This is an important limitation for our analysis: in the inference routine we will undertake in the next Section, we want to be able to explore the full parameter space and consider models for which this range of masses (or even higher) plays a significant role. For this reason, we fit the cross-correlation terms with two key objectives: reducing the noise associated with the poor statistics at the high mass end of the halo mass function, and providing a means to sensibly extrapolate 
the behavior of the cross-correlation functions up to $\log_{10} M/\msun = 14.0$ ($\log_{10} M/\msun = 14.5$) at $z=4$ ($z=2.5$). This extrapolation allows us to recover well-behaved posterior distributions (see Sec. \ref{sec:results}) that provide a complete description of the different models described by our parameters. Its validity and the associated caveats are discussed in detail in Sec. \ref{sec:discussion_caveats}

We provide details on the fitting of the cross-correlation terms $\xi_h(M_j, M_k; r)$ in Appendix \ref{sec:appendix_fitting}. In short, we divide all the cross-correlation terms, $\xi_h(M_j, M_k; r)$, by a reference correlation function,  $\xi_{\rm ref}(r)$, and fit the results with a 3-d polynomial to capture the residual dependencies on the two masses and on radius.
In the rest of this Section, we show the results of the fits for the auto-correlation functions in different mass bins at $z=4$ (Figure \ref{fig:simulations}, right panel; the same plot for $z=2.5$ is shown in Appendix \ref{sec:appendix_fitting_z2.5}). In other words, we plot the correlation functions for bins of equal mass, $\xi_h(M_j, M_j; r)$, together with the fits that are meant to reproduce these functions, $\xi_{h, {\rm fit}}(M_j, M_j; r)$ (gray lines)\footnote{The global fits to all the cross-correlation terms $\xi_h(M_j, M_k; r)$ at both redshifts are shown in Appendix \ref{sec:appendix_fitting}.}. 
Lower mass bins correspond to lower values of the auto-correlation functions, and vice-versa. 

We assign errors to the $\xi_h(M_j, M_j; r)$ points based on the Poissonian statistics of the pair counts; note that in this way we are underestimating the real uncertainties on the data points because we are not including the effects of cosmic variance and of other sources of systematics. For this reason, when assessing the robustness of our fits, it makes little sense to discuss them in terms of statistical errors. We therefore compare the simulated data and the model fits in terms of relative differences between the two (lower right panel of Figure \ref{fig:simulations}). 
These differences are generally at the level of $\lesssim10\%$ for all bins but the highest one (i.e., $\log_{10} M/\msun = 13.25-13.5$), which is easily recognizable because it has the largest Poissonian uncertainties. As already mentioned before, at very large masses correlation measurements from simulations become noisy (and thus unreliable) due to the small number of halos in the snapshots. Even in this extreme case, however, the fit provides a satisfactory description of the shape and normalization of the correlation function in the simulations, with a relative difference that is still smaller than the uncertainties on the \citeS07 observed data (which are at the level of $50-100\,\%$; see Sec. \ref{sec:obs_data}). 

Using dashed grey lines, we also plot in Fig. \ref{fig:simulations} the auto-correlations functions for the two bins $\log_{10} M/\msun = 13.5-13.75$ and $\log_{10} M/\msun = 13.75-14.0$, as obtained by extrapolating our fitting functions to masses higher than the ones probed by the simulations. We see that the trend of the auto-correlation functions with halo mass is well preserved by these extrapolations; further discussion on this can be found in Sec. \ref{sec:discussion_caveats} and Appendix \ref{sec:appendix_fitting_z2.5}. 

Finally, we note that relative differences between our fits and the values of correlation functions extracted from simulations tend to be larger at very large scales ($r\gtrsim100\,\cMpc$). This is also due to the fact that simulation-based values become less reliable in this regime. There are two reasons for that: first, the finite size of the box reduces the number of very large-scale pairs that are available. 
Secondly, at $r\gtrsim100\,\cMpc$ the behavior of correlation functions becomes non-trivial due to the presence of the baryon acoustic oscillations (BAO) peak. This is especially difficult to model given the very coarse radial bins we have chosen. Due to these limitations of our model, we simply exclude scales larger than $r\gtrsim100\,\cMpc$ from our analysis.

\section{Data-model comparison} \label{sec:model_data}

In the previous Section, we have described how to obtain the two observables of interest (i.e., the QLF and the quasar auto-correlation function) starting from a CLF and a simulation-based analytical description of the halo mass function and of the halo cross-correlation functions. We now provide more details on the actual comparison between our model and observational data.

\subsection{Overview of observational data}
\label{sec:obs_data}

We start by giving a brief description of the data that we compare the model with. 
Our main goal is to explain the very strong quasar clustering measured by \citeS07 at $z\approx4$. Thus, we make use of the \citeS07 data for the projected auto-correlation function ($w_\mathrm{p}/r_\mathrm{p}$). Note that the authors assume that the data points are independent (because the quality of the data is not good enough to extract a covariance matrix), so we will do the same and use the \citeS07 errors assuming that the covariance matrix for the data is diagonal. 
We use the ``good fields'' data (see \citeS07 for the definition) as they are supposed to be more reliable and -- since they show stronger clustering -- have proven to be the hardest to reproduce theoretically \citep[e.g.,][]{shankar_2010}. As already mentioned, we exclude the data at very large scales ($r>100\,\cMpc$) from our analysis because they are particularly challenging to measure both in observations \citep[e.g.,][]{eftekharzadeh2015} and in simulations (see the end of the last Section).

In the subsequent analysis, we are also interested in reproducing the quasar clustering at lower redshift. For this purpose, we use data from the Baryon Oscillation Spectroscopic Survey (BOSS, \citealt{eftekharzadeh2015}; hereafter, \citeE15). We focus on the redshift range $z=2.2-2.8$, where the majority of the BOSS quasars reside. We use the data for the projected correlation function, $w_\mathrm{p}(r_\mathrm{p})$, in the radial range $4\,\cMpch < r_\mathrm{p} < 25\,\cMpch$. In this region, the \citeE15 data are considered more reliable by the authors and an estimate for the error covariance matrix is available. 

One of the key points of our analysis is that, while the QLF includes all quasars known in a given redshift bin, the quasar auto-correlation function is usually measured by considering only quasars above a given luminosity threshold $L_{\rm thr}$. This is an important point to take into account in our model (see eq. \ref{eq:qhmf}-\ref{eq:hod}), as the presence of such a threshold may bias the inferred clustering significantly. The flux limit employed for the \citeS07 measurements is $m_i=20.2$ (where $m_i$ is the apparent magnitude in the $i$ band). In order to convert this to a value of $L_{\rm thr}$, we first convert the apparent magnitude $m_i$ to an absolute magnitude, $M_{1450}$, using the $K(z)$ correction\footnote{The conversion between $m_i$ and $M_{1450}$ can be made using $K_{i,1450}(z)$, which is defined as: $M_{1450}(z) = m_i - 5\log_{10} \left(d_\mathrm{L}(z)/{\rm Mpc}\right) -25-K_{i,1450}(z)$, with $d_L(z)$ being the luminosity distance at redshift $z$. Following \citet{lusso2015}, we set $K_{i,1450}(z=4)\approx -1.9$.} (see, e.g., \citealt{kulkarni2019} and references therein). We obtain that $m_i=20.2$ corresponds to $M_{1450} = -25.72$ at $z=4$. We then convert this value to a bolometric luminosity by applying the bolometric corrections provided by \citet{runnoe2012}\footnote{The bolometric correction for $\lambda=1450$ \r{A} 
is $\log_{10} L_{\rm iso}/\ergs  = 4.745 + 0.910 \log_{10} \lambda L_\lambda/\ergs$. $L_{\rm iso}$ is the bolometric luminosity calculated under the assumption of isotropy, and it is related to the real bolometric luminosity $L$ via a factor that accounts for the viewing angle, $L = 0.75 \, L_{\rm iso}$}. We get a value for the \citeS07 luminosity threshold equal to $\log_{10} L_{\rm thr}/\ergs = 46.7$.  

As for the \citeE15 clustering data at $z\approx2.5$, the luminosity threshold that we should employ is more subtle. While the authors consider the entirety of the BOSS sample \citep[][]{ross2013}{}{} for their clustering analysis, they also show that this sample is highly incomplete at low luminosities. This is an issue in the context of our model, as, when setting a threshold $L_{\rm thr}$, we are implicitly assuming that the sample is complete above the threshold. Given that properly modeling completeness in the \citeE15 sample is outside the scope of this work, we set the value of $L_{\rm thr}$ to the $25$th percentile of the luminosity distribution of the observed quasars at $z=2.5$. This value represents a compromise between taking into account part of the highly incomplete sample of faint quasars that are included in the clustering analysis and minimizing the bias that these quasars generate in the predicted clustering. By considering Figure 3 in \citeE15, we set this threshold value to a $M_i(z=2)$ magnitude of $-25.3$. Following \citet{lusso2015}, we convert this to $M_{1450} = M_i(z=2)+1.28=-24.02$, and finally to a bolometric threshold of $\log_{10} L_{\rm thr}/\ergs = 46.1$.

As for
the QLF, there are many different estimates available. For the sake of consistency with the clustering measurements, we choose to employ the UV-bright quasar catalogue compiled by \citet[][hereafter \citeK19]{kulkarni2019}. These authors provide a homogenised catalogue of 80,000 color-selected AGN from
redshift $z=0$ to $7.5$, together with MCMC-based estimates of the QLF at all redshifts. We employ this dataset and select quasars at different redshifts according to our models. For the model at $z=4$, we set $3.5<z<4.5$ (largely consistent with the \citeS07 high-z sample); in this range, the bright end of the QLF is determined by the same SDSS quasars that are used to compute the clustering \citep[][]{schneider2010}, whereas the low-luminosity quasars are presented in \citet{glikman2011}. The model at $z=2.5$, instead, is entirely determined by quasars observed by the BOSS survey \citep[][]{ross2013}.

Quasars in the \citeK19 dataset are binned according to their $M_{1450}$ magnitude, and the uncertainties are computed using Poisson statistics. However, as also discussed in \citeK19, the QLF data always present significant systematic errors due to, e.g.,
uncertainties in the quasar selection. This implies that the quoted uncertainties on the QLF data may be significantly underestimated, as is also evident from the large scatter (up to $\approx1\,{\rm dex}$) between different estimates of the QLF that are available in the literature \citep[e.g.,][]{shen2020_bol_qlf,grazian2023}{}{}. These issues are particularly problematic in our framework, as our goal is to perform statistical inference by simultaneously matching the quasar luminosity and auto-correlation functions, and that can only be done properly if the associated uncertainties are well understood and treated. Therefore, in order to avoid biases in our inference analysis owing to very small formal statistical uncertainties on the QLF, we add a systematic error to every QLF measurement in quadrature to the Poisson ones determined by \citeK19. That is, the uncertainties on our QLF data points are set to be $\sigma^2 = \sigma^2_{\rm sys} + \sigma^2_{\rm count} $, where $\sigma^2_{\rm sys}$ ($\sigma^2_{\rm count} $) stands for the systematic (statistical) uncertainty. We adopt a constant systematic uncertainty of $0.2~\mathrm{dex}$
for the $z\approx 4$ dataset and of $0.05~\mathrm{dex}$ for the $z\approx2.5$ one. This implies a systematic relative uncertainty of $\approx45\%$ ($\approx10\%$) for $z=4$ ($z=2.5$). These values are chosen to be similar to the average relative statistical uncertainties at the two redshifts considered ($\approx40\%$ and $\approx5\%$ at $z=4$ and $z=2.5$, respectively).

As the final step, we convert the values of the quasars' absolute magnitudes in \citeK19, $M_{1450}$, to bolometric luminosities using the \citet{runnoe2012} bolometric corrections. We stress the fact that our results are independent of the adopted bolometric corrections, as our model could easily be expressed in terms of quasars' UV magnitudes only. However, as discussed in Sec. \ref{sec:methods_basic}, we choose to convert everything to bolometric luminosities for consistency with previous work on the subject.

\subsection{Likelihood functions}
\label{sec:bayesian_framework}

We employ a Bayesian framework to write the posterior distributions for our model parameters. As described in Sec. \ref{sec:methods_basic}, the model consists of a log-normal CLF centered on a power-law dependence of the quasar luminosity on halo mass.
The free parameters are the normalization and slope of the quasar luminosity-halo mass relation ($L_{\rm ref}$ and $\gamma$, respectively), the logarithmic scatter around this relation ($\sigma$), and the fraction of quasars that are active at any given moment ($f_{\rm on}$). The final set of parameters, $\Theta$, is then: $(\sigma, L_{\rm ref}, \gamma, f_{\rm on})$.

We set flat priors on $\sigma$ and $\gamma$, and flat priors on the logarithm of $L_{\rm ref}$ and $f_{\rm on}$ (see e.g. \citealt{jeffreys1946}). Our priors span the following parameter ranges: 
$\sigma\in\left(0.1\,{\rm dex}, 1.5\,{\rm dex}\right)$; $\log_{10} L_{\rm ref}/\ergs\in\left(44.0,46.6 \right)$; $\gamma\in\left(0.5,3 \right)$; 
$\log_{10} f_{\rm on}\in\left(-3,0\right)$. 
These limits are chosen in order to focus on the region of the parameter space where models are physically motivated (e.g., the scatter in the $L_c-M$ relation is unlikely to be smaller than $0.1~\mathrm{dex}$).

In what follows, we want to fit the QLF and the auto-correlation function both independently and simultaneously. We can get 
constraints on these two observables by setting the following likelihood functions:
\begin{equation}
    \mathcal{L}^{(k)}({\bf d}^{(k)}\,|\Theta) = \frac{1}{(2\pi)^{n/2}|\Sigma|^{1/2}} \exp\left(-\frac{1}{2}(\mathbf{y}-\boldsymbol\mu)^\top \Sigma^{-1}(\mathbf{y}-\boldsymbol\mu)\right), \label{eq:likelihood}
\end{equation}
where $k\in\{{\rm QLF},\,{\rm corr}\}$ stands either for the quasar luminosity function data or for the auto-correlation function data. As for the other variables, ${\bf d}^{(k)}$ stands for the set of $n$ data points with means $\mathbf{y}$ and covariance $\Sigma$ coming from observations, whereas $\boldsymbol\mu$ stands for the set of values predicted by our models. 

With the above likelihood, the results for the correlation function (``corr") are found to not be very constraining, as there is a large set of models that produce the correct clustering but substantially under(over)-estimate the number density of bright quasars. Therefore, when quoting results for the correlation function only, we provide an additional integral constraint by imposing that the model matches the observed number density of bright quasars. We integrate the QLF above the luminosity limit used for the clustering measurements (see Sec. \ref{sec:obs_data}), and obtain an estimate for the number density of bright quasars, $n_{\rm bright}$. The associated uncertainty, $\sigma_{\rm bright}$, is determined by using different realizations of the QLF fits from \citeK19. 
Then, we predict the number of quasars with a luminosity above this threshold, $L_{\rm thr}$, based on our model ($n_{\rm model}$),
and use the following likelihood: 
\begin{equation}
    \mathcal{L}^\mathrm{(corr+nden)} = \frac{\exp^{-(n_{\rm bright} - n_{\rm model})^2/\sigma_{\rm bright}^2 }}{\sqrt{2\pi}\sigma_{\rm bright}} \,\mathcal{L}^\mathrm{(corr)}. 
\end{equation}
Note that we do not fit to the shape of the QLF, but only to the total abundance of quasars above $L_{\rm thr}$. This is an integral constraint that favors models producing a physically reasonable total number of bright quasars.

Finally, we provide joint constraints on the parameters by fitting the QLF and the auto-correlation function simultaneously. In other words, we write the joint likelihood distribution as the product of the two likelihoods (we assume that the two measurements are independent, and weigh the two dataset equally):
\begin{equation}
    \mathcal{L}^\mathrm{(joint)} = \mathcal{L}^\mathrm{(QLF)}\,\mathcal{L}^\mathrm{(corr)}.
    \label{eq:likelihood_joint} 
\end{equation}
Note that for the joint likelihood distribution, we consider $\mathcal{L}^\mathrm{(corr)}$ (rather than $\mathcal{L}^\mathrm{(corr+nden)}$), as the QLF already provides an implicit constraint on the total abundance of luminous sources. 

\section{Results} \label{sec:results}

In this section, we describe the results we obtain by fitting our model to the observed quasar luminosity and auto-correlation functions, both independently (``QLF'' and ``corr+nden'' cases) and simultaneously (``joint'' case; see Sec. \ref{sec:bayesian_framework} for the definitions). Henceforth, we will refer to the ``QLF'' model as ``QLF only'' in order to distinguish our model from the QLF itself. We first consider the $z=4$ case -- which is the main focus of this paper -- and then discuss the results at lower redshift ($z=2.5$) as well. 

\begin{table}
\centering
\setlength{\extrarowheight}{3pt}
\caption{Best-fitting parameter values for our model-data comparison at $z=4$ (see the main text for definitions of the different parameters, as well as eq. \ref{eq:likelihood}-\ref{eq:likelihood_joint}). ``QLF only'' refers to the quasar luminosity function data only, ``corr+nden'' refers to the auto-correlation function data in conjunction with the number density of bright quasars, and ``joint'' refers to the combined QLF+auto-correlation function data. The last column shows the minimum value of the normalized chi-squared (see text for details).}
\begin{tabular}{ c | c c c c | c }
\toprule
 Quantity & $\sigma$ & $\log_{10} L_{\rm ref}$ [$\ergs$]    & $\gamma$   & $f_{\rm on}$ [\%] & $\chi^2_{\rm norm}$\\ 
\midrule
QLF only  &  
$0.38$ &
$46.4$     &   
$0.78$ &
$0.2$ &
$2.2/5$ 
\\
corr+nden    &   
$0.10$ &
$44.4$     &   
$2.99$ &
$100$ &
$4.6/4$\\
joint &
$0.11$ &
$45.1$     &   
$2.07$ &
$66$ &
$12.9/12$\\
 \bottomrule
\end{tabular}
\label{tab:best_fitting}
\end{table}

\begin{figure*}
	\centering
	\includegraphics[width=\textwidth]{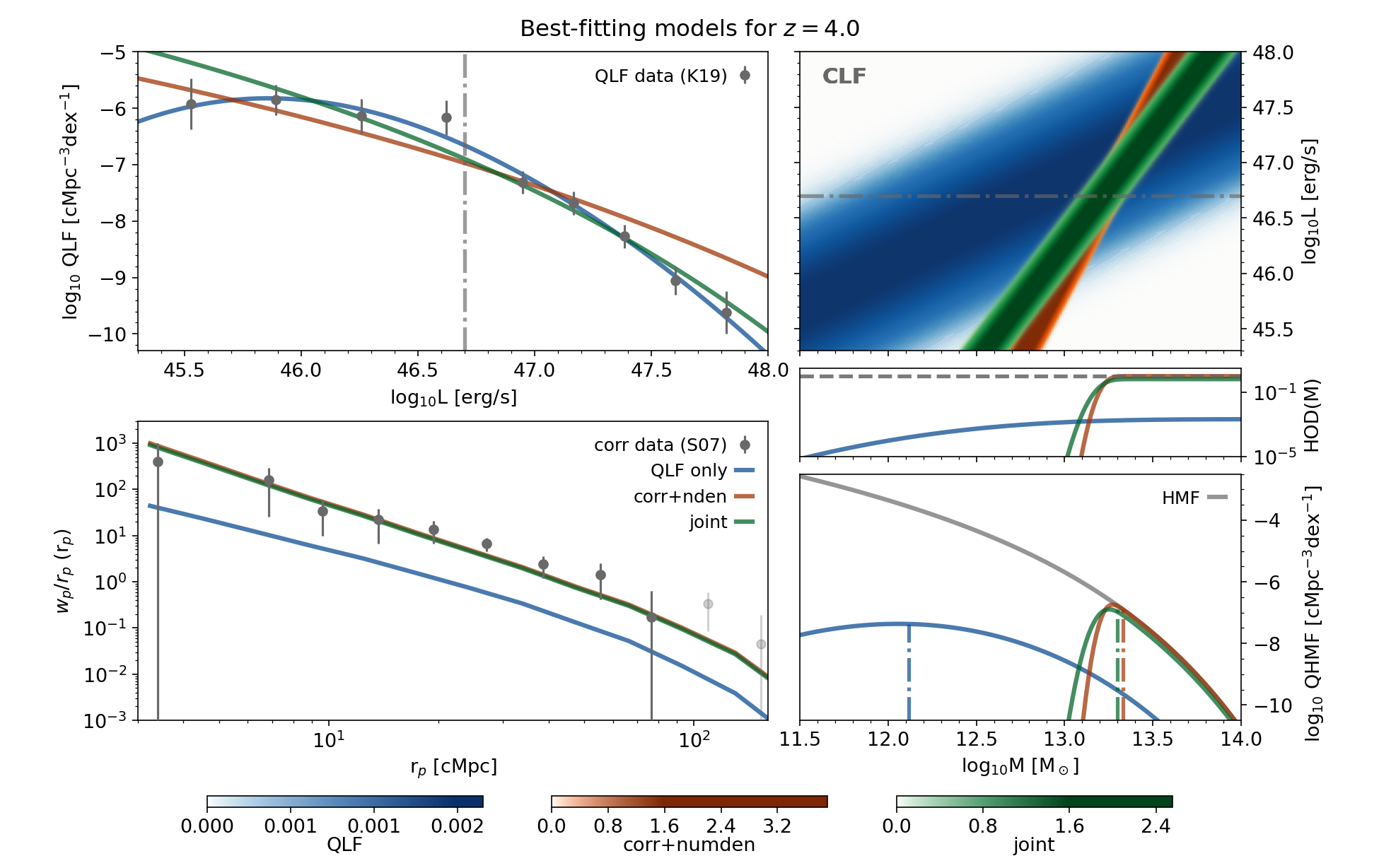}

	\caption{Overview of our model-data comparison at $z=4$. The blue, orange, and green colors refer to the best-fitting models for the ``QLF only'', ``corr+nden'', and ``joint'' likelihood distributions, respectively (see text for details and Table \ref{tab:best_fitting} for the parameters' values). The upper right panel shows the Conditional Luminosity Function (CLF$(L|M)$), with the associated color bars at the bottom representing the probability density. The horizontal dot-dashed gray line in the same panel (and the vertical one in the upper left panel) refers to the luminosity threshold for clustering measurements, $\log_{10} L_{\rm thr}$. Integrating the CLF along the luminosity axis above this threshold gives the Halo Occupation Distribution (HOD; middle right panel), which can be combined with the Halo Mass Function (gray line in the lower right panel), to give the Quasar-Host Mass Function (QHMF; coloured lines in the lower right panel). Vertical dot-dashed lines in the lower right panel are the median values of the QHMF distribution, $M_{\rm med}$ (see eq. \ref{eq:qhmf_med}). The QHMF is then used to predict the projected quasar auto-correlation function, $w_\mathrm{p}(r_\mathrm{p})$ (lower left panel). The $z\approx4$, \citeS07 data for the auto-correlation function are also shown in the same panel (data outside the fitting range, $3<r_\mathrm{p}/\cMpc<100$, are shown as semi-transparent points). The upper left panel shows our predictions for the Quasar Luminosity Function (QLF), together with the $z\approx4$ data from \citeK19. 
  \label{fig:results_overview}
 	}
\end{figure*}

\subsection{Analysis at $z\approx4$}
\label{sec:results_z4}

 \begin{figure*}
	\centering
	\includegraphics[width=0.48\textwidth]{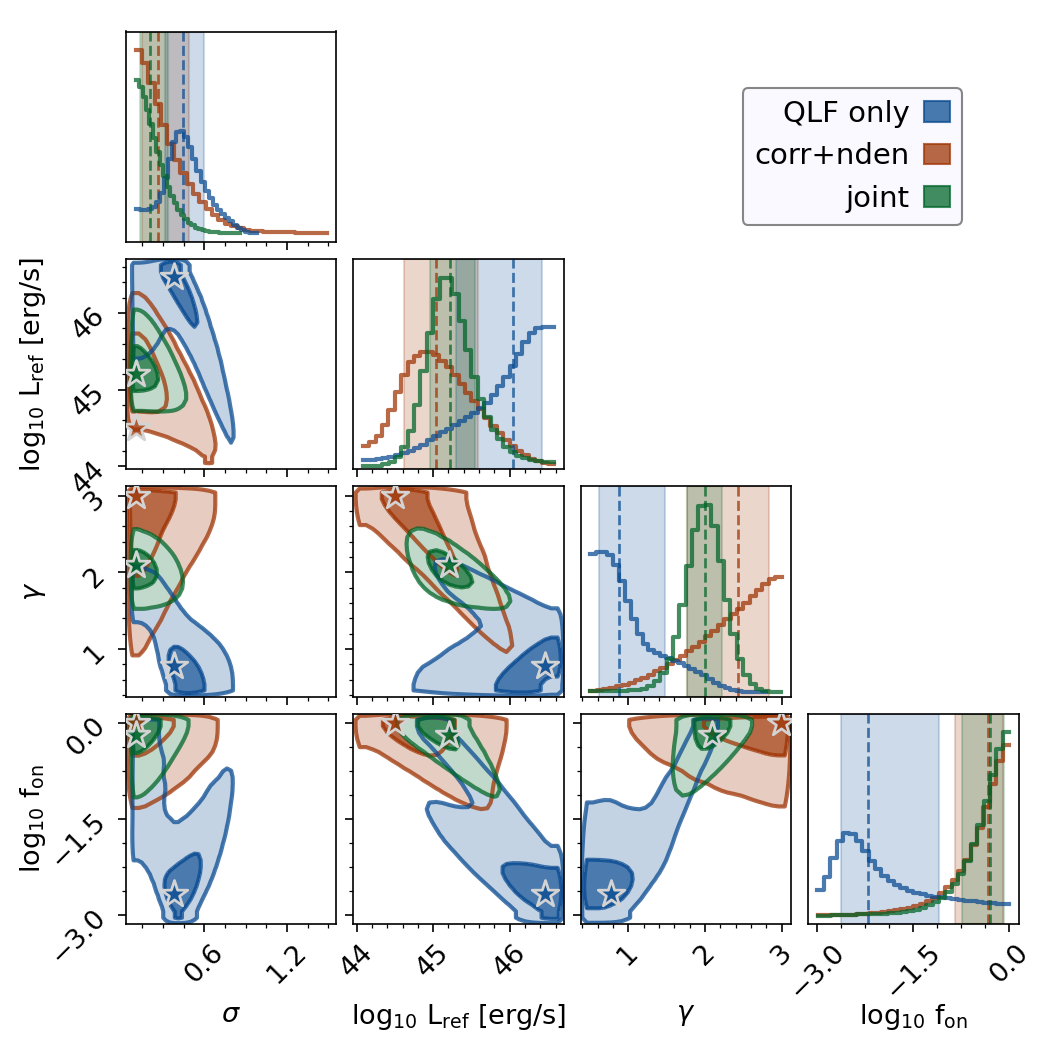}
 	\includegraphics[width=0.48\textwidth]{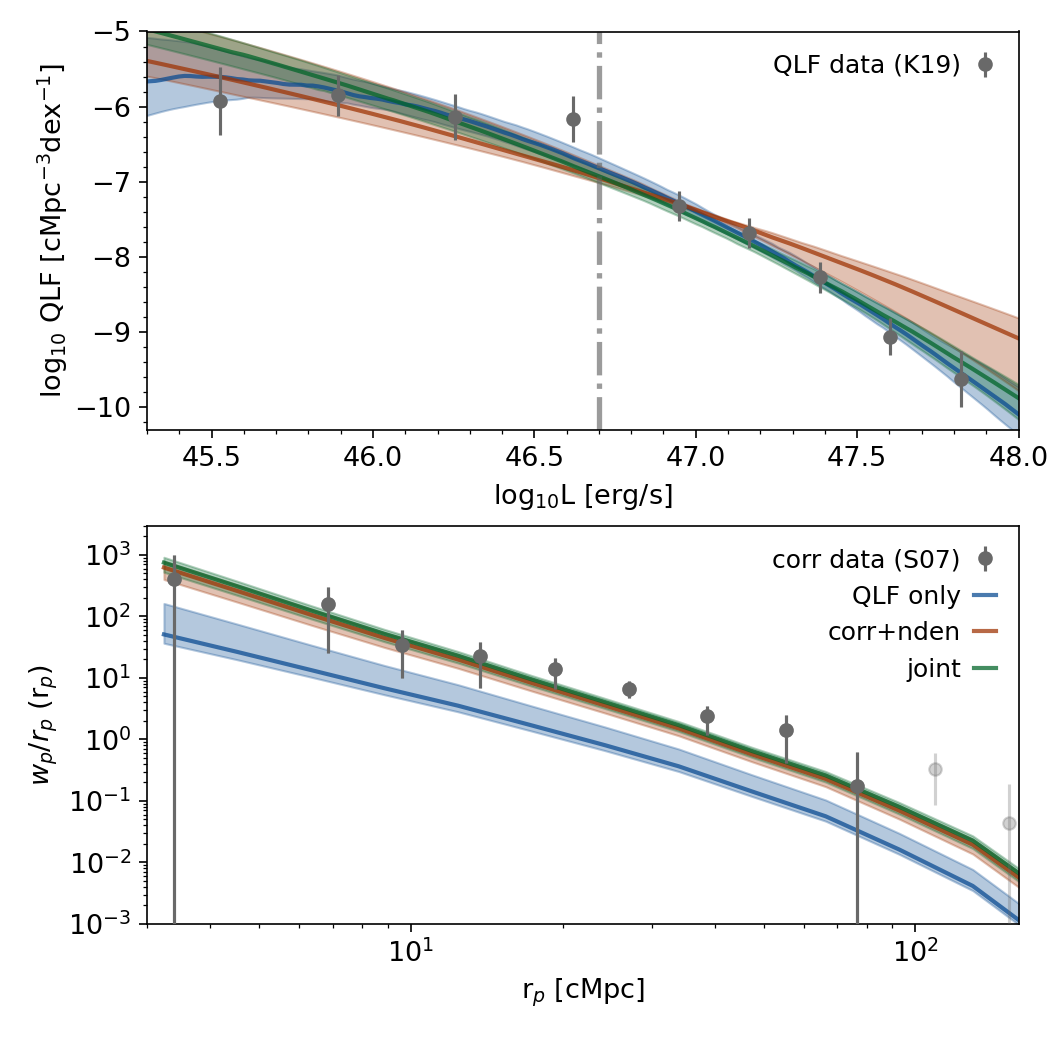}
	 \caption{\textit{Left:}
	Corner plots of the 4-d posterior distributions for the different cases described in Sec. \ref{sec:results_z4} (blue for the ``QLF'' model, orange for ``corr+nden'', and green for ``joint''). 
	Contours in the 2-d histograms highlight the $1\sigma$ and $2\sigma$ regions, whereas the dashed lines in the 1-d histograms represent the median values of the parameters (with $1\sigma$ errors shown as shaded regions).
    Best-fitting parameters from Table \ref{tab:best_fitting} (see also Fig. \ref{fig:results_overview}) are shown with star symbols in each corner plot. 
	\textit{Right}: Comparison of the predicted quasar luminosity (top) and auto-correlation (bottom) functions with the observational data from \citeK19 and \citeS07, respectively. The color coding is the same as in the left panel. Median values (solid lines) and $1\sigma$ uncertainty regions (shaded areas) are obtained by randomly sampling the Markov chains for the posterior distribution $100$ times. Data points for the auto-correlation function that are outside of our fitting range (see Sec. \ref{sec:obs_data}) are shown as semi-transparent points in the bottom right panel. The vertical dot-dashed line in the upper right panel is the luminosity threshold for quasar clustering, $L_{\rm thr}$ (see Sec. \ref{sec:model_data}).
  \label{fig:results_mcmc}
	}
\end{figure*}

As a first step, we are interested to know whether our model can reproduce the two observables. We can answer this question by employing a simple optimization algorithm to find the maximum of the likelihood distributions (or, equivalently, of the posterior distributions) for the three cases of interest: quasar luminosity function only (``QLF only''), correlation function + number density of bright quasars (``corr+nden''), and quasar luminosity and correlation functions together (``joint''). The maxima of the likelihoods represent our best-fitting models, which we can then compare directly with observations (see Section \ref{sec:results_mcmc} for the results of the full parameter inference). 

In Table \ref{tab:best_fitting}, we report these best-fitting parameters for the cases mentioned above. 
Figure \ref{fig:results_overview} shows our model predictions at the maximum likelihood
parameter values for the CLF, the HOD, the QHMF, the QLF, and the projected quasar autocorrelation function ($w_\mathrm{p}/r_\mathrm{p}$); see Fig. \ref{fig:method_summary} for a schematic overview of these quantities.  

In the top right panel of Fig. \ref{fig:results_overview}, we show the conditional luminosity functions, ${\rm CLF}(L|M)$, as a function of the quasar luminosity $L$ and the halo mass $M$. The three cases ``QLF only'', ``corr+nden'', and ``joint'' are shown with different colors (blue, orange, and green, respectively). The associated color bars at the bottom of the Figure represent the probability densities for the different CLF cases. 
Integrating the CLF above the luminosity threshold $L_{\rm thr}$ (gray dashed-dotted line in the CLF panel), we obtain the halo occupation distribution (HOD; middle right panel; eq. \ref{eq:hod}). 
Combining the HOD with the halo mass function (HMF), we get the Quasar-Host Mass Function (QHMF; eq. \ref{eq:qhmf}); this is shown in the bottom right panel, together with the $z=4$ HMF (gray line). The two left panels show the predictions for the observable quantities: the auto-correlation function is shown on the bottom left, together with data from \citeS07; the quasar luminosity function (eq. \ref{eq:qlf}) is shown on top (data are from \citeK19). While the auto-correlation function is obtained from the QHMF via eq. \ref{eq:quasar_corr_func}-\ref{eq:projected_corrfunc}, the QLF is the result of integrating along the mass axis of the CLF weighted by the HMF (eq. \ref{eq:qlf}).

Overall, looking at the two left panels of Figure \ref{fig:results_overview}, we conclude that in all cases the models constitute very good fits to the data they are meant to reproduce (see below for the caveat on the ``QLF only'' case). In order to quantify this, we use reduced chi-squared statistics, $\chi^2_{\rm norm}=\chi^2/\nu_{\rm ndof}$, where $\nu_{\rm ndof}$ is the number of degrees of freedom (i.e., the number of data points minus the number of parameters). We find $\chi^2_{\rm norm}=2.2/5$, $\chi^2_{\rm norm}=4.6/4$, and $\chi^2_{\rm norm}=12.9/12$ for the ``QLF only'', ``corr+nden'', and ``joint'' cases, respectively. These values are also shown in Table \ref{tab:best_fitting} for reference.   

One striking feature of the best-fitting models is that they have very different properties, as can be seen in the top right panel of Fig. \ref{fig:results_overview} and the best-fitting parameters shown in Table \ref{tab:best_fitting}. All of them are characterized by low values of the scatter in the quasar luminosity-halo mass relation, $\sigma$, but the offset, slope, and normalization of this relation vary significantly between the models. 

The ``QLF only'' model 
shows an approximately linear relation with a high value of the reference luminosity $L_{\rm ref}$. 
As a result, the characteristic mass of halos hosting quasars with a luminosity above $L_{\rm thr}$ is low ($\log_{10} M/\msun \approx 12.35$, lower right panel of Fig. \ref{fig:results_overview}). This has two consequences. Firstly, halos with $\log_{10} M/\msun \approx 12-12.5$ are much more abundant than the number of observed quasars, and thus a very low active fraction ($f_{\rm on} \approx 0.1\%$) is needed to match the QLF. Secondly, such a low characteristic mass for the halos hosting luminous quasars implies a low value for the quasar auto-correlation function, in conflict with the \citeS07 measurements (lower left panel). In fact, we see that the best-fitting model for the ``QLF only'' case
does not fare well when compared with the clustering data. 

The ``corr+nden'' model, instead, finds a much larger characteristic host mass for bright quasars ($\log_{10} M /\msun \approx 13-13.5$). Such a large mass is achieved by packing quasars in almost all the most massive halos. This is done thanks to a few key ingredients (upper right panel of Fig. \ref{fig:results_overview}): a low value of the quasar luminosity at the reference mass of $\log_{10} M_{\rm ref}/\msun =12.5$, a highly non-linear relation between quasar luminosity and halo mass ($\gamma\approx3$) and a very low scatter in this relation $\sigma\approx0.1$. The first two parameters determine the mass, $\log_{10} \Tilde{M}$, at which the quasar luminosity -- halo mass relation crosses the luminosity limit $L_{\rm thr}$. The second and third parameters, instead, determine how sharply the HOD drops at masses lower than $\log_{10} \Tilde{M}$ (middle right panel of Fig. \ref{fig:results_overview}). 
The extreme scenario implied by our best-fitting model is needed to reproduce the measured auto-correlation function. Indeed, the shape and normalization of the \citeS07 data are very well reproduced by our model (Fig. \ref{fig:results_overview}, lower panel) at the scales considered in the analysis ($3 \,\cMpc \lesssim r_\mathrm{p}\lesssim 100\,\cMpc$).

Besides fitting the auto-correlation function, the ``corr+nden'' model aims to reproduce the number density of quasars above the luminosity threshold $\log_{10} L_{\rm thr}$. This is also achieved by the best-fitting model, which predicts a number density $n_{\rm model}=3.18\times10^{-8}\,\cMpc^{-3}$, $0.5$ standard deviations higher than the observational value of $n_{\rm bright}=2.73\times10^{-8}\,\cMpc^{-3}$. The shape of the QLF, however, is not well reproduced by the model, because it overpredicts the abundance of very bright systems and underpredicts the abundance of $\log_{10} L/\ergs \approx 46-47$ quasars. This is due to the fact, despite the very low value of $\sigma$, the strong non-linearity in the quasar luminosity-halo mass relation ($\gamma\approx3$) associates a large fraction of the massive halos to the brightest observable quasars.  

When we simultaneously fit both the quasar auto-correlation and the luminosity function (``joint'' model), we obtain results that are quite similar to the ``corr+nden'' case, and are compatible with the same extreme scenario in which quasars are packed in the most massive halos, i.e., a non-linear quasar luminosity-halo mass relation with a steep slope and very small scatter, low value of the quasar luminosity at the reference mass, and a large active fraction of quasars. The quasar luminosity-halo mass relation for the ``joint'' model is however not as extreme 
as the one for the ``corr+nden'' model, as it is characterized by a lower value of the power-law exponent, $\gamma\approx2$. This has very little impact on the auto-correlation function, as the quasar-host mass functions (lower right panel of Fig. \ref{fig:results_overview}) are very similar in the two cases. It does have an effect, however, on the shape of the QLF, with the ``joint'' model providing a better fit, especially at the very bright end. 

Overall, the QLF is very well reproduced by the ``joint'' model, with the exception of the low-luminosity end ($\log L/\ergs\approx45.5$). In this region, the largest differences between the ``QLF only'' and the ``joint'' model appear, with the ``QLF only'' model faring better at predicting a flattening of the shape of the QLF. This flattening, however, is an artificial feature of our model, originating from the prior assumption that halos with a mass lower than $\log_{10} M/\msun =11.5$ do not host quasars. We consider this issue not worthy of further investigation, as the faint-end of the QLF is still largely unconstrained by data, and deeper observations are needed to probe its behavior at the high redshift \citep[e.g.,][]{akiyama2018, parsa2018, giallongo2019, harikane2023, grazian2023}. Furthermore, our primary focus here is to interpret the bright quasars that are also probed by clustering surveys. It is possible that a more flexible quasar luminosity-halo mass relation 
is necessary to account for the abundance of low-luminosity systems.

\subsubsection{MCMC analysis}
\label{sec:results_mcmc}

Given that our models are a good representation of the observational data, we can proceed further with inference and determine how well the data constrain the model parameters. We explore the posterior distributions using a Markov-Chain Monte Carlo (MCMC) approach. We employ the Python package \code{emcee} \citep{foreman2013emcee} to sample the posteriors using the affine-invariant ensemble prescription \citep{goodman_prescription}. We place $m=48$ walkers distributed randomly in the parameter space and evolve them for $N>10^5$ steps. We set the final number of steps so that our chains are at least 100 times longer than the auto-correlation time $\tau$ \citep[see e.g.,][]{sharma2017markov}, and thin the chains considering only one element every $\tau$ steps in order to account for auto-correlations. We also discard the first $10^3$ elements of every chain to account for the burn-in phase.

Figure \ref{fig:results_mcmc} (left panel) shows the corner plot for the 4-d posterior distributions (as a function of $\sigma, L_{\rm ref}, \gamma, f_{\rm on}$) for the three cases considered in the analysis (``QLF'', ``corr+nden'', and ``joint'').  The best-fitting model for each of these cases, which was discussed above and shown in Fig. \ref{fig:results_overview}, is highlighted with a star symbol in the corner plots.
The samples of the posterior distributions obtained by the Markov Chains are then used  
to obtain predictions for the quasar luminosity and the auto-correlation functions; we compare these quantities with the data in the right panels of Figure \ref{fig:results_mcmc}. 

As expected, the ``QLF only'' and ``corr+nden'' models peak in very different regions of the parameter space. The ``corr+nden'' model constrains the parameters to the region with $\sigma\lesssim 0.5$, $\gamma\gtrsim 2$, $\log_{10} L_{\rm ref}/\ergs \approx44.5-45.5$, and $f_{\rm on }$ close to unity. This region of the parameter space is the only one that is compatible with the above-mentioned scenario in which bright quasars are active only in the most massive halos. This is also the reason why there are no models predicting stronger quasar clustering than observed (Figure \ref{fig:results_mcmc}, right panel), as our models are already predicting the strongest possible clustering compatible with the observed abundance of bright systems.

The ``QLF only'' model, on the other hand, peaks at lower $\gamma$ and $f_{\rm on}$, larger $\log_{10} L_{\rm ref}$, and a value of $\sigma$ which is larger than the ``corr+nden'' but still moderately low ($\sigma \approx 0.3-0.5$). However, the distribution for the ``QLF only'' case is much more complex, and therefore the resulting constraints on the parameters are not as straightforward. In particular, there is a region of the parameter space that is well within the constraints given by the auto-correlation function, and for which the ``QLF only'' model also has a good match with 
the QLF data (at the $\lesssim 2\sigma$ level). Unsurprisingly, this is the region where the ``joint'' posterior distribution is located (green contours in Fig. \ref{fig:results_mcmc}). 

The reason why this region of parameter space can reproduce both the QLF and the auto-correlation function can be understood as follows.  
As mentioned in the introduction, the behavior of the QLF at the bright end is very different from the one of the HMF, with the latter being characterized by an exponential cutoff that is not present in the QLF. 
This is usually explained by assuming that the more abundant population of lower-mass halos can also contribute to the population of luminous quasars \citep[see e.g.,][]{ren_trenti2020}. Indeed, this is what our fiducial ``QLF only'' model seems to suggest (see also Fig. \ref{fig:results_overview}), as the very high quasar luminosity predicted for the $10^{12.5}\,\msun$ halo population implies that even with $\approx0.4$ dex of scatter the correct shape of the luminosity function can be reproduced. In this picture, quasars are relatively common phenomena arising in the bulk of the halo population at that redshift, with a very low duty cycle of $\varepsilon_{\rm DC}\approx0.1\%$. 
However, even a scenario in which only the most massive halos are active as bright quasars (with a duty cycle $\varepsilon_{\rm DC}\gtrsim50\%$) can be compatible with the observed shape of the QLF. In this second case, the non-linearity in the quasar luminosity--halo mass relation plays a key role 
in mapping the exponential cutoff of the HMF into the power-law bright end slope of the QLF, while at the same time packing bright quasars only in the most massive hosts. While both of these scenarios provide a good description of the QLF -- differing significantly only at lower luminosities -- only the latter is also compatible with a very large clustering length of quasars.

In conclusion, despite the fact that the quasar luminosity and auto-correlation functions alone provide relatively loose constraints on the shape of the Conditional Luminosity Function (CLF), when considered in conjunction they are able to determine a very well-defined region in the parameter space for which a good agreement with all observational data is achieved (right panel of Fig. \ref{fig:results_mcmc}). This is the most significant conclusion of our analysis, and we will discuss it further in Sec \ref{sec:discussion}.

\subsection{Comparison with $z\approx2.5$}
\label{sec:results_lowz}

 \begin{figure*}
	\centering
	\includegraphics[width=0.48\textwidth]{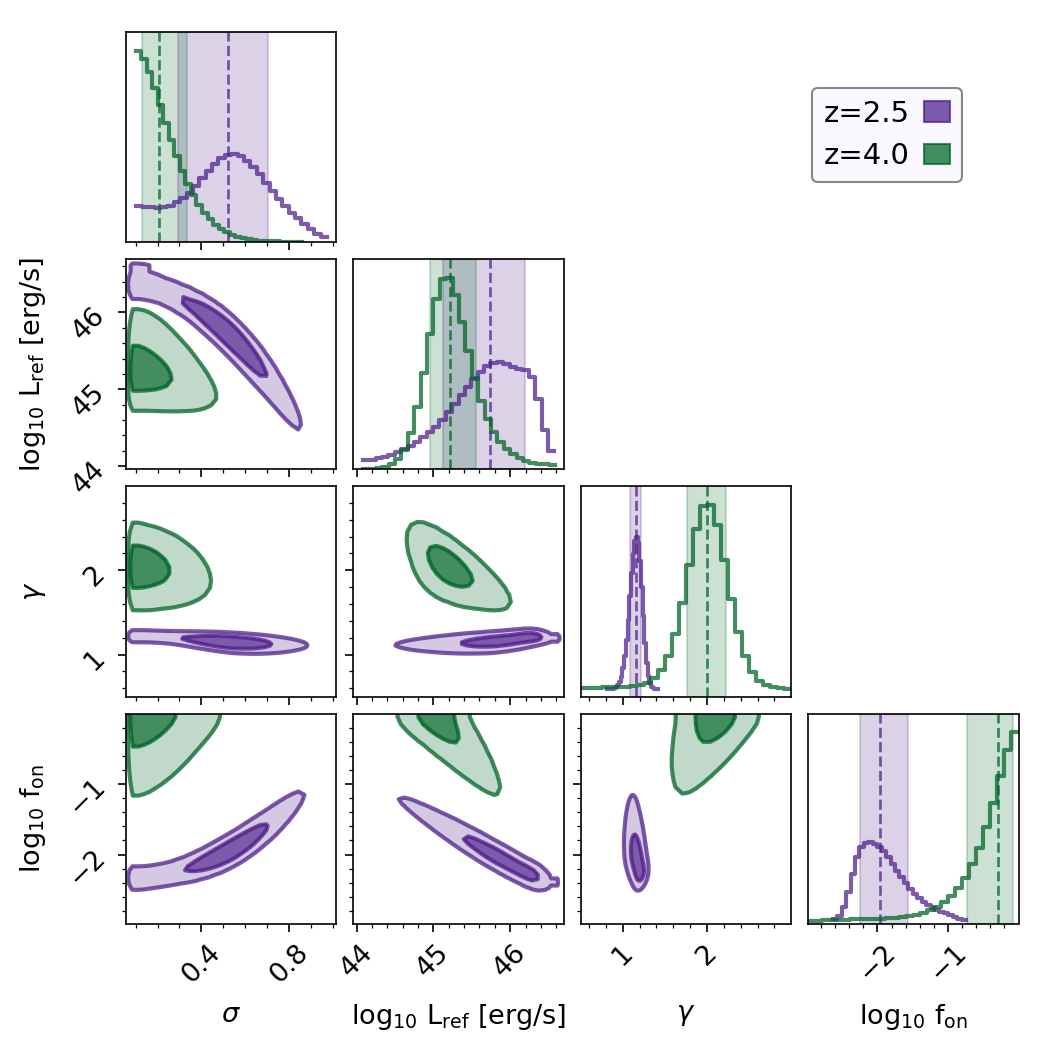}
 	\includegraphics[width=0.48\textwidth]{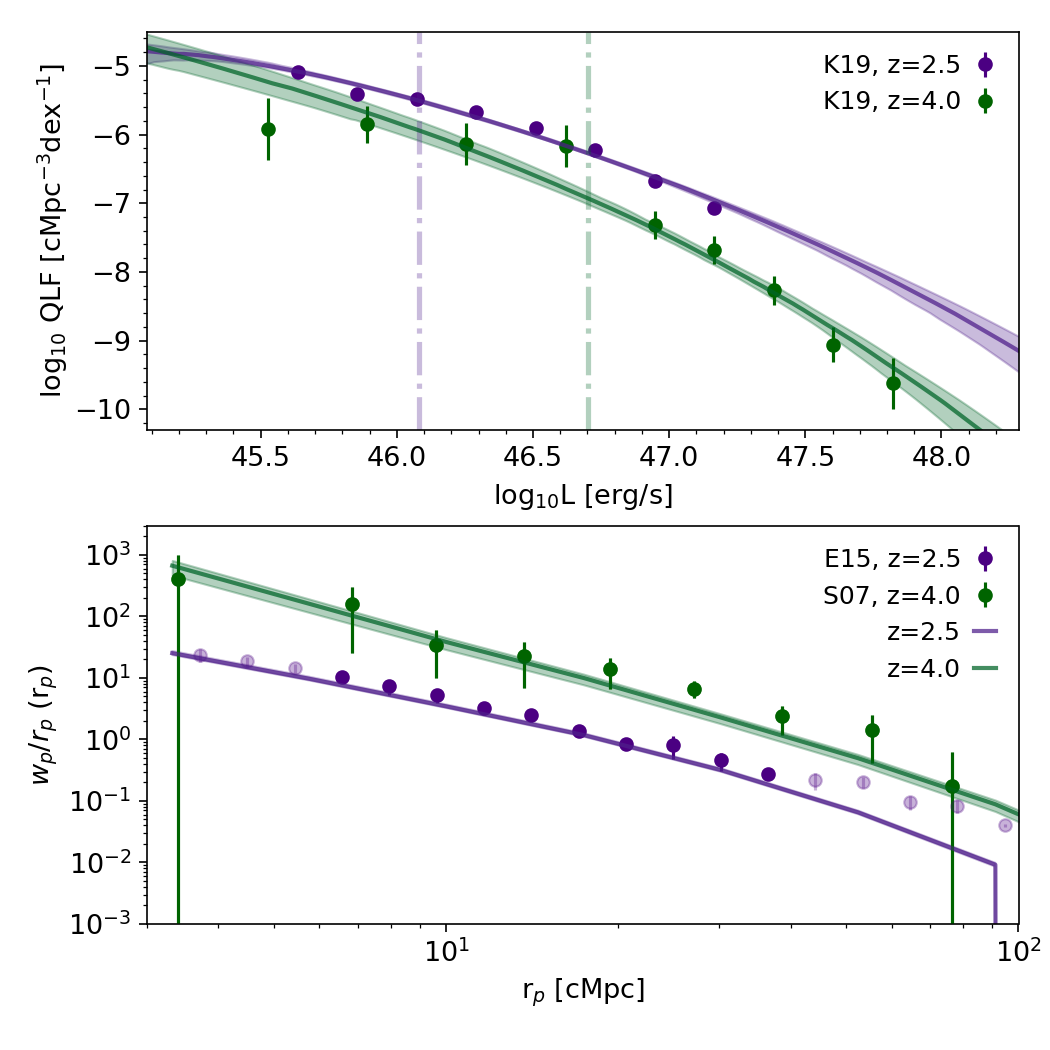}

  \caption{Same as Fig. \ref{fig:results_mcmc}, but for different redshifts ($z=4$ in green and $z=2.5$ in purple). The results always refer to the ``joint'' model (Sec. \ref{sec:results_z4}). The vertical dot-dashed lines in the upper right panel are the luminosity thresholds, $L_{\rm thr}$, used to measure quasar clustering at the two redshifts. Data points for the auto-correlation function that are outside of our fitting range because they are considered not reliable (see Sec. \ref{sec:obs_data}) are shown with semi-transparent colours in the bottom right panel. 
  \label{fig:results_mcmc_redshift}
 	}
\end{figure*}

Having applied our model to $z\approx4$ data, it is also important to test whether the model is flexible enough to reproduce observations at lower redshifts, where the observed strength of quasar clustering is not as extreme. We note that our goal in this paper is not to provide a complete and self-consistent evolutionary description of quasar properties across cosmic time, but simply to strengthen the conclusions we have drawn in the previous section by showing that the same framework can also be applied to describe the spatial and luminosity distributions of quasars at different epochs. In particular, we focus on the redshift range $z=2.2-2.8$, where the 
the BOSS survey \citep[][]{ross2013, eftekharzadeh2015} has provided solid measurements of the quasar luminosity and auto-correlation functions. We choose this data set because it is sufficiently different from the one at $z\approx4$ to suggest that the properties of quasars may have varied significantly in a relatively short amount of time. 

For simplicity, we focus here on the ``joint'' models only. In other words, we run the MCMC-based algorithm with the same setup as in Sec. \ref{sec:results_mcmc} fitting the quasar luminosity function and the auto-correlation function simultaneously. 
Figure \ref{fig:results_mcmc_redshift} shows the resulting corner plots for the posterior distribution of the ``joint'' model at $z=2.5$ (purple), together with the one at $z=4.0$ (green; same as Fig. \ref{fig:results_mcmc}) for comparison. We report the values of the resulting 1-d constraints on the model parameters for both redshifts in Table \ref{tab:mcmc}. In the right panel of Figure \ref{fig:results_mcmc_redshift}, we show the predictions of our models based on randomly sampling the Markov chains for the posterior distributions, together with the data that we aim to reproduce. We note that, as mentioned in Sec. \ref{sec:obs_data}, we only include the data points for the \citeE15 auto-correlation function at $z\approx2.5$ in the range $6~\cMpc\lesssim r_\mathrm{p}\lesssim 40~\cMpch$. This is because data outside this range are not considered reliable and not included in the covariance matrix estimation (see \citeE15). Indeed, we find that our model provides a good match to the \citeE15 data within the fitting range, but it is significantly lower than the measured data at larger scales. Given the strong biases that may be associated with large-scale estimates of the correlation function, we do not consider this issue worthy of further investigation.  

The corner plots in Figure \ref{fig:results_mcmc_redshift} show that the regions of the parameter space constrained by the two redshifts are quite different. Interestingly, the shape of the $z=2.5$ posterior distribution exhibits non-trivial behavior in the 2-d projections, yielding tight constraints on the $\gamma$ parameter, but also strong degeneracies between $\sigma$, $\log_{10} L_{\rm ref}$, and $f_{\rm on}$. 
In general, however, the different parameters are well constrained, even better than at $z=4$ due to the higher sensitivity of the data. The resulting 1-d posteriors for $z=2.5$ and $z=4$ peak at a similar value of $\log_{10} L_\mathrm{ref}$, but they are quite different for the other parameters. Lower-$z$ results are characterized by a lower value of $\gamma$ ($\approx1.15$) and $f_\mathrm{on}$ ($\approx0.01$), and a higher value of the scatter in the $L-M$ relation, $\sigma$.

The top panel of Figure \ref{fig:discussion_redshift} shows how these posteriors translate into distributions for the QHMFs (eq. \ref{eq:qhmf}). In this plot, the QHMFs for $z=2.5$ and $z=4$ are shown, together with the HMFs at the same redshifts (semi-transparent lines). Uncertainties on the QHMFs are computed by randomly subsampling the Markov chains for the posterior distributions. 
The two QHMFs are quite different, reflecting the differences in the level of clustering measured at the two redshifts. In the $z=4$ case, quasars only reside in the most massive systems ($\log_{10} M/\msun\gtrsim 13$), with the QHMF distribution tightly following the HMF (see also Fig. \ref{fig:results_overview} for the best-fitting model). At $z=2.5$, instead, the QHMF distribution has a lower median value ($\log_{10} M /\msun\approx 12.5$) and it is much broader, with a large range of halos of different masses capable of hosting quasars. 

The differences in the QHMF translate directly into different measurements for the quasar duty cycle. As discussed in Sec. \ref{sec:methods_basic}, we define the quasar duty cycle (eq. \ref{eq:duty_cycle}) as the ratio between the QHMF integrated above the median value of its distribution, $M_{\rm med}$ (eq. \ref{eq:qhmf_med}), and the HMF integrated above the same threshold. For $z=2.5$, we find a value of the duty cycle equal to $\varepsilon_{\rm DC}=0.4\pm0.1\,\%$, whereas for $z=4$ we find $\varepsilon_{\rm DC}=33^{+34}_{-23}\,\%$. We note that these values are closely related to the values of the $f_{\rm on}$ parameter (Table \ref{tab:mcmc}), which describes the active fraction of quasars at any given moment. Only in the case of a perfectly deterministic $L-M$ relation (i.e., with zero scatter), however, would we find a duty cycle exactly equal to $f_{\rm on}$. In the presence of scatter in the $L-M$ relation, the shape of the QHMF can vary significantly with respect to the one of the HMF, and this changes the fraction of quasars that are above the threshold luminosity, $L_{\rm thr}$, at any given mass, and hence the quasar duty cycle.

However, we should mention the caveat that these results are obtained by setting two different luminosity thresholds, $L_{\rm thr}$, at the two redshifts considered, according to the minimum luminosities imposed in the respective clustering measurements. As shown in the top right panel of Figure \ref{fig:results_mcmc_redshift}, the $z=2.5$ luminosity threshold is $\approx0.6$ dex lower than the one at $z=4$ ($L_{\rm thr}=46.1\,\ergs$ and $L_{\rm thr}=46.7\,\ergs$ at $z=2.5$ and $z=4$, respectively).
Changing the value of $L_{\rm thr}$ may have direct consequences for the QHMF, HOD, and quasar duty cycle, since all these quantities have an explicit dependence on $L_{\rm thr}$ (eq. \ref{eq:qhmf}-\ref{eq:duty_cycle}). 

In order to provide a fair comparison between these quantities at the two redshifts considered in the analysis, we impose the same $L_{\rm thr}$ at both redshifts by using the $z=4$ luminosity threshold (i.e., $L_{\rm thr}=46.7\,\ergs$) to recompute the above-mentioned quantities at $z=2.5$. While the duty cycle remains unchanged (within uncertainties), we find that although the QHMF is still very broad, its normalization and median value are lower and higher, respectively. In particular, the median value of the QHMF, $M_{\rm med}$ (eq. \ref{eq:qhmf_med}) shifts from $\log_{10} M_{\rm med}/\msun\approx12.5$ to $\log_{10} M_{\rm med}/\msun\approx12.8$. This suggests a mild dependence of clustering on luminosity, as more luminous quasars tend to be hosted by more massive halos. However, given that the QHMF distribution is very broad in both cases, there is a strong overlap between the populations of very bright ($\log_{10} L /\ergs\gtrsim46.5$) and moderately luminous ($\log_{10} L /\ergs\approx 46-46.5$) quasars in terms of their host halo masses. 

We leave a detailed analysis of the implications of our model in terms of the luminosity dependence of quasar clustering for future work. Here, we simply note that even when adopting the same luminosity threshold, we find a remarkable difference between $z=4$ and $z=2.5$ quasars. The former are very extreme objects, hosted only by the most massive halos that are present at that redshift, representing $4-5\sigma$ peaks in Gaussian random fields (\citealt{colossus_diemer2018}; see also Sec. \ref{sec:introduction}). The latter, instead, are hosted by much more common halos at $z=2.5$, which are only slightly over-massive with respect to the bulk of the halo population at that redshift ($2-3\sigma$ peaks). For this reason, despite the increase in the quasar number density between $z=4$ and $z=2.5$, the quasar duty cycle -- which measures how abundant quasars are with respect to their host population -- decreases by two orders of magnitudes between the same two redshifts.

In conclusion, our data-model comparison reveals that the same parametrization of the CLF employed at $z=4$ is also able to reproduce the data at lower-$z$, with a significant evolution of the CLF parameters reflecting a remarkable change in the physical properties of quasars with cosmic time. In the following, we further discuss the implications of these findings.

\begin{table}
\centering
\setlength{\extrarowheight}{3pt}
\caption{Constraints on the model parameters based on the corner plots shown in Figure \ref{fig:results_mcmc_redshift}.}
\begin{tabular}{c | c c c c }
\toprule
 Redshift & $\sigma$ & $\log_{10} L_{\rm ref}$ [$\ergs$]    & $\gamma$   & $f_{\rm on}$ [\%]\\ 
\midrule
$z=2.5$ &
$0.52^{+0.18}_{-0.22}$ &
$45.7^{+0.46}_{-0.61}$     &   
$1.15^{+0.06}_{-0.07}$ &
$1.01^{+1.59}_{-0.04}$ \\
\midrule
$z=4$ &
$0.20^{+0.13}_{-0.08}$ &
$45.2^{+0.3}_{-0.3}$     &   
$2.00^{+0.22}_{-0.23}$ &
$51^{+32}_{-31}$ \\
\bottomrule
\end{tabular}
\label{tab:mcmc}
\end{table}

\section{Discussion} \label{sec:discussion}

In the analysis performed above, we could successfully match the quasar luminosity and auto-correlation functions at two different redshifts provided that: (a) there exists a non-linear relation between quasar luminosity and halo mass, and the non-linearity increases with redshift; (b) the scatter in this relation is fairly small ($\sigma \lesssim 0.3-0.6$) and decreases significantly with redshift; (c) in accordance with this relation, luminous quasars ($\log_{10} L/\ergs \gtrsim 46.5$) are hosted by halos with mass $\log_{10} M/\msun \approx 13-13.5$ ($\log_{10} M/\msun \approx 12.5-13$) at $z=4$ ($z=2.5$); (d) the quasar duty cycle is a strong function of redshift, with a very low $\varepsilon_{\rm DC}\approx0.4\%$ at low-$z$ that increases to $\varepsilon_{\rm DC} \approx 30\%$ at $z=4$. In the following, we further elaborate on this picture by investigating its implications for SMBH accretion and growth 
and by placing it in the context of previous work on the subject. We end the section by highlighting the main strengths and weaknesses of our analysis.

\subsection{Implications for quasars' physical properties}

\subsubsection{Black hole mass and accretion efficiency}

The CLF posits an empirical relation between quasar luminosity and halo mass. However, many quasar population models \citep[e.g.,][]{conroy_white2013, veale2014, trinity}
built this relation on physical grounds by relating the quasar luminosity to the mass of the central black hole, and this latter mass to the mass of either the host halo or the host
galaxy/bulge. We can relate these two approaches by introducing the Eddington ratio $\eta$, which is defined by the following relation:
\begin{equation}
    L =\zeta \,\eta\, M_{\rm BH},\label{eq:eta_mbh}
\end{equation}
where $M_{\rm BH}$ is the mass of the black hole, and $\zeta=3.67\times 10^4\,\lsun/\msun$ is a constant factor. 

Then, we assume, e.g., that the mass of a black hole is determined solely by the mass of the host halo. In other words, we introduce a probability $P(M_{\rm BH}|M)$ for the mass of the black hole given the halo mass. If we also write the ``Eddington ratio distribution'' ERDF$(\eta| M_{\rm BH}, M)$ in terms of the other quantities considered, the conditional luminosity function reads:
\begin{equation}
    {\rm CLF}(L|M) = \int \frac{\d M_{\rm BH}}{\xi M_{\rm BH}}  \,{\rm ERDF}\left(\frac{L}{\xi M_{\rm BH}} \middle| M_{\rm BH}, M\right)    
    P(M_{\rm BH}|M).  \label{eq:clf_erdf_mbh}
\end{equation}
In this way, we have related the CLF -- which is an empirically determined stochastic relationship between quasar luminosity and halo mass --  
to two other distribution functions 
(the ERDF and the black hole mass distribution) that have a clear physical meaning, being related to the physics of black hole accretion and growth. 

In order to make this relationship explicit in our analysis, we can simply rewrite the quasar luminosity as the product of the Eddington ratio and the black hole mass (eq. \ref{eq:eta_mbh}). In this way, we can explicitly study how these two parameters -- albeit completely degenerate -- depend on the mass of the host halo, $M$, according to our model. The middle panel of Figure \ref{fig:discussion_redshift} illustrates this dependence. In this panel, we employ the ${\rm CLF}(L|M)$ relation given by our model to write the probability distribution for the product of the Eddington ratio and the black hole mass-halo mass ratio, $P(\eta M_{\rm BH}/M |M)$. Note that we divide the product $\eta M_{\rm BH}= L/\zeta$ by the halo mass, $M$, because we expect black hole mass and halo mass to be approximately proportional based on local scaling relations \citep[e.g.,][]{efstathiou_rees1998,white_2008, booth_schaye2010, 
marasco2021}{}{} and because we can then work with a dimensionless quantity. 
Redshifts in the middle panel of Figure \ref{fig:discussion_redshift} are color-coded as in the top panel and in Figure \ref{fig:results_mcmc_redshift}. Median values and uncertainties for $P(\eta M_{\rm BH}/M |M)$ are extracted by randomly sampling the Markov chains for the posterior 
distributions, as well as the $L-M$ relation given by our model (see the caption for details). 

While at $z=2.5$ the median value of $\eta M_{\rm BH}/M$ shows only a weak trend with halo mass, the situation is much different at $z=4$, with the product $\eta M_{\rm BH}/M$ strongly correlating with $M$. This can be achieved by assuming that either the black hole mass, the Eddington ratio, or both increase with halo mass. In other words, for very massive hosts black holes are either particularly massive (with a black hole mass-halo mass ratio higher than for lower-mass counterparts) or efficiently accreting (i.e., with large Eddington ratios). This trend is driven by the fact that the measured strong clustering at $z\approx4$ requires that the most luminous quasar population is completely dominated by high-mass hosts. 

It is also useful to cast these constraints in terms of galaxy stellar masses. This can be done by exploiting one of the parameterizations of the halo mass-stellar mass 
relation that are available in the literature. Here, we use the redshift-dependent 
halo mass-stellar mass relation from \citet{behroozi2013} to rewrite $P(\eta M_{\rm BH}/M |M)$ in terms of the galaxy mass $M_*$, i.e., $P(\eta M_{\rm BH}/M_* |M)$. For simplicity, we neglect the scatter between stellar mass and halo mass in this conversion. The bottom panel of Figure \ref{fig:discussion_redshift} shows how the product of the Eddington ratio and the black hole mass-galaxy mass ratio varies as a function of halo mass. This is especially interesting in light of the fact that there is long-standing evidence in favor of a linear (or quasi-linear) relation between black hole and galaxy masses in the local universe
(the so-called $M_{\rm BH}-M_*$ relation, see e.g., \citealt{magorrian_1998, kormendy_ho2013, reines_volonteri2015}). In the same panel (Figure \ref{fig:discussion_redshift}), we plot with a dashed red line the expectation 
for the product $\eta M_{\rm BH}/M_*$ as a function of $M$, based on assuming the local $M_{\rm BH}-M_*$ relation as measured by \citet{reines_volonteri2015}, converting galaxy masses to halo masses according to \citet{behroozi2013}, and setting a fixed Eddington ratio of $\eta=1$. 
The scatter around this quantity (red shaded region) only considers the scatter in the $M_{\rm BH}-M_*$ relation as quoted by \citet{reines_volonteri2015}. Due to the fact that the $M_{\rm BH}-M_*$ relation is almost linear, the product between the Eddington ratio and the black hole mass-galaxy mass ratio is almost independent of halo mass, with an average constant value of $\eta M_{\rm BH}/M_*\approx-3.5$.

Comparing this expectation based on local relations to the predictions of our model for $z=2.5,4$, we find that for low halo masses ($\log_{10} M/\msun \lesssim 12.5$) the predictions tend to agree within the uncertainties (see below for the caveat on extrapolating below $\log_{10} M/\msun \approx 12$). For larger masses, however, the difference between local predictions and our models becomes quite significant. At $z=2.5$, there is a mild but significant positive trend of increasing $\eta M_{\rm BH}/M_*$ with $M$. This trend becomes even steeper and tighter at $z=4$, with the product $\eta M_{\rm BH}/M_*$ ranging from $\approx-3.5$ for $\log_{10} M/\msun \lesssim 12.5$ to $\approx-2$ for $\log_{10} M/\msun \lesssim 13.5$. 
The trend can be interpreted considering that the galaxy formation efficiency at $z=2-4$ peaks at halo masses around $\log_{10} M_{\rm peak}/\msun \approx 12-12.5$; as a consequence, the stellar mass-halo mass relation flattens at masses higher than $\log_{10} M_{\rm peak}$ \citep[][]{behroozi2013,universe_machine}{}{}. Our model, on the other hand, does not predict any flattening in the quasar luminosity-halo mass relation for masses $\log_{10} M>\log_{10} M_{\rm peak}$. Hence, the product $\eta M_{\rm BH}/M_*$ becomes a steep function of halo mass for $\log_{10} M /\msun \gtrsim 12.5$.
We believe the absence of a flattening in the quasar luminosity-halo mass relation at the high mass end is not simply a consequence of the chosen parametrization for the CLF. By experimenting with different parameterizations of the CLF, we found that any departure from a steep, non-linear relation between quasar luminosity and halo mass is incompatible with the measured value of the clustering (especially at $z=4$), as a flattening of this relation would lower the characteristic halo mass of bright-quasar hosts.  Therefore, we conclude that while galaxy growth appears to be quenched at the high mass end, even at high redshifts \citep[e.g.,][]{universe_machine}{}{}, this does not seem to be the case for black hole growth, as black holes in very massive halos need to be very massive and/or accreting efficiently. Indeed, observational evidence for an evolution of the $M_{\rm BH}-M_*$ relation has been found repeatedly at high-$z$ (implying over-massive black holes) together with signs of an increase in the median value of the ERDF with redshift (e.g., \citealt{vestergaard_osmer2009,wu2022,maiolino2023,pacucci2023, stone2023}; however, see \citealt{li2022}; \citealt{trinityII} for a discussion of selection biases).

We conclude by noting that, in our analysis, the shape of the CLF is actually constrained by data only in a limited range of halo masses. For low halo masses, the corresponding quasar luminosities fall in a range where quasar clustering has never been measured and estimates for the QLF are not available (or highly uncertain). On the other hand, for very high halo masses (and hence very high quasar luminosities), quasars become so rare that estimates for the QLF are once again very uncertain. Moreover, if the quasars are luminous enough to be completely above the luminosity threshold for clustering, then the exact behavior of the luminosity as a function of halo mass becomes irrelevant. 
Therefore, in all panels of Figure \ref{fig:discussion_redshift}, we show the regions in halo mass where our constraints on the CLF are based purely on extrapolations as dotted lines. This mainly concerns low halo masses ($\log_{10} M/\msun \lesssim 12.5$) at $z=4$, and both very low ($\log_{10} M/\msun \lesssim 12$) and very high ($\log_{10} M/\msun \gtrsim 13.5$) masses at $z=2.5$.

\subsubsection{Quasar lifetime and the growth of high-$z$ black holes}

While the empirical relation between quasar luminosity and halo mass gives valuable information on the connection between black holes, their accretion efficiency, and their host halos/galaxies, another key piece of the puzzle resides in the inferred values of the quasar duty cycle, $\varepsilon_\mathrm{DC}$. This quantity is defined as the fraction of halos that are hosting bright quasars at any given time (see Sec. \ref{sec:methods_basic}). If quasar activity is a stochastic process, however, the duty cycle is also equal to the total fraction of time in which a black hole is active as a bright quasar during the lifetime of an average host halo. In other words, the duty cycle is an average constraint on the total lifetime of a quasar, $t_\mathrm{Q}$. Following \citet{martini2001} (see also \citealt{martini_2004}, \citealt{haiman_hui2001}), we can simply assume that the characteristic lifetime of a halo is roughly equal to the age of the universe $t_\mathrm{U}(z)$, and get an estimate for the quasar lifetime by writing $t_\mathrm{Q}=t_\mathrm{U}(z)~\varepsilon_\mathrm{DC}$. Using the values of $\varepsilon_\mathrm{DC}$ obtained in Sec. \ref{sec:results_lowz}, we get $t_\mathrm{Q}\approx 0.1-1~\mathrm{Gyr}$ at $z\approx4$, and $t_\mathrm{Q}\approx 10-15~\mathrm{Myr}$ at $z\approx2.5$. 

\begin{figure}
	\centering
	\includegraphics[width=0.5\textwidth]{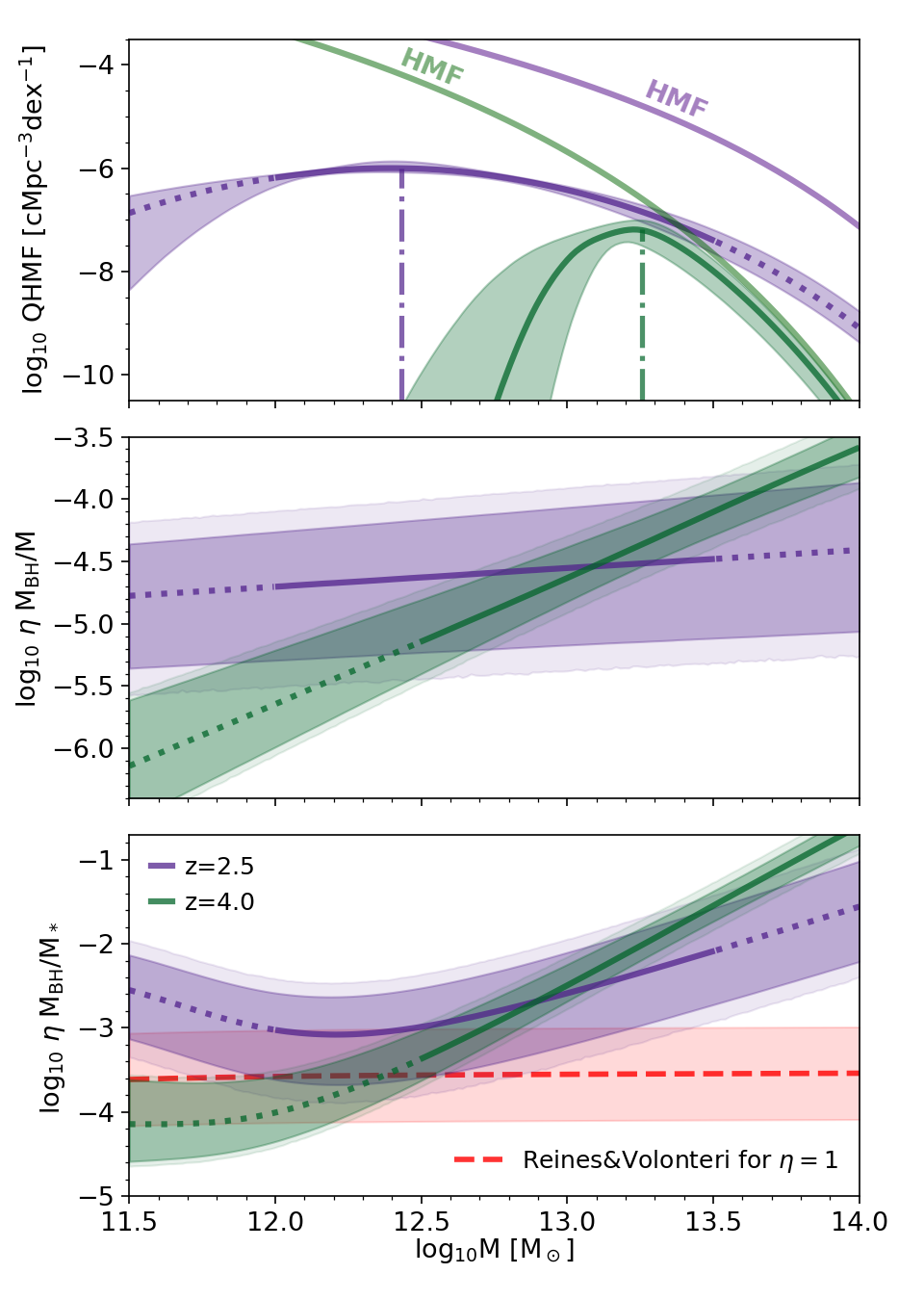}

	 \caption{{\it Top:} Quasar-host mass function (QHMF) at $z=2.5$ (solid purple line) and $z=4$ (green), according to our model. These functions and their respective uncertainties are the median and the 16th and 84th percentiles of the distributions obtained by randomly subsampling the Markov chains of the posteriors shown in Figure \ref{fig:results_mcmc_redshift}. The halo mass functions (HMFs) for both redshifts are shown with semi-transparent lines, whereas the dashed-dotted lines indicate the median values of the QHMF distributions. In all panels, regions in the halo mass spectrum where the behavior of the conditional luminosity function (CLF) is purely extrapolated and not explicitly constrained by data are shown with dotted lines. {\it Middle:} Same as the top panel, but showing the dependence on halo mass of the product between the Eddington ratio ($\eta$) and the black hole-halo mass ratio ($M_{\rm BH}/M$). In this case, there are two sources of scatter: the uncertainty on the model given by the posterior distribution and the intrinsic scatter coming from the $\sigma$ parameter in the CLF. We plot the former with a darker shading, whereas the total contribution of the two sources of scatter is shown with a lighter shading. {\it Bottom:} Same as the middle panel, but showing the quantity $\eta \, M_{\rm BH}/M_*$ instead (with $M_*$ being the galaxy mass). The relation between halo mass and galaxy mass is taken from \citet{behroozi2013}. The red dashed line shows the prediction for $\eta \, M_{\rm BH}/M_*$ assuming the \citet{reines_volonteri2015} relation between black hole and galaxy masses (with the shading showing the scatter in the relation), and setting $\eta=1$.
  \label{fig:discussion_redshift}
 	}
\end{figure}

The quasar lifetime is one of the most fundamental quantities for understanding the role that SMBHs play in a cosmological context. According to the standard picture of SMBH growth \citep[e.g.,][]{lynden_bell1969}, luminous quasars are powered by gas accretion onto a SMBH, and the
rest mass energy of this material is divided between the small fraction ($\approx 10\%$) of radiation that we observe, and the growth of the black hole. In this picture, a phase of luminous quasar activity translates directly into a buildup of mass for the central SMBH. This provides a direct connection between the total luminosity emitted by quasars over cosmic time and the total mass residing in SMBHs in the local Universe (the so-called ``Soltan argument'', \citealt{soltan1982}).
If the quasar lifetime is long compared to the Hubble timescale (i.e., the duty cycle is large), then the buildup of the total SMBH mass has taken place in only a small fraction of host galaxies that were active as bright quasars for a large fraction of their lifetimes. A short quasar lifetime, on the other hand, implies that most galaxies have undergone a brief bright quasar phase during their evolution history. The results of our analysis suggest that the latter scenario is valid at cosmic noon ($z\approx1-3$), when most of the SMBH growth has taken place \citep[e.g.,][]{shen2020_bol_qlf}{}{}. The short quasar lifetime we find at $z\approx2.5$ is, in fact, a direct consequence of the fact that quasar activity at that redshift takes place in relatively common halos with a broad distribution of host halo masses (top panel of Figure \ref{fig:discussion_redshift}). 
Opposite conclusions can be obtained by considering our $z\approx4$ results. In this case, we find that the large duty cycle translates into a quasar lifetime that is a large fraction of the Hubble time ($t_\mathrm{Q}\approx0.1-1~\mathrm{Gyr}$). This implies that SMBH growth may be radically different in the young Universe as compared to cosmic noon. As suggested by the $z\approx4$ QHMF in Figure \ref{fig:discussion_redshift} (top panel), quasar activity at high $z$ takes place only in the few most massive halos that are present at that redshift, and hence these systems are active as bright quasars for a large fraction of cosmic time.

Estimating the quasar lifetime at high $z$ is even more compelling in light of the fact that observations of very massive black holes powering luminous quasars at $z\gtrsim5$ challenge our standard paradigm for black hole formation and growth \citep{mazzucchelli2017,farina2023}. In the standard picture, black holes follow an Eddington-limited exponential growth with a timescale that is equal to the ``Salpeter time'', $t_\mathrm{S}\approx40~\mathrm{Myr}$. At high $z$, models suggest that there is just enough cosmic time to grow the observed SMBH masses starting from massive seeds of $\approx10^3-10^5\,\msun$ \citep[e.g.,][]{inayoshi2020}{}{}. For this reason, gauging the quasar lifetime is important because it offers an indirect probe of whether sustained accretion on SMBHs can take place at high $z$ in the form of bright quasar activity. The long lifetime we infer at $z\approx4$ is indeed consistent with this picture, providing an argument in support of models for Eddington-limited growth of high-$z$ black holes. We can provide a rough estimate for this argument by considering as a characteristic host halo mass the median value of the $z\approx4$ QHMF (Figure \ref{fig:discussion_redshift}, top panel), $\log_{10} M_\mathrm{med}/\msun\approx13.3$. 
If we assume accretion at the Eddington rate ($\eta=1$), we can translate this characteristic halo mass into a black hole mass using the relation between $\eta \,M_\mathrm{BH}/M$ and $M$ (middle panel of Figure \ref{fig:discussion_redshift}): we get $\log_{10} M_\mathrm{BH}/\msun\approx9$ \citep[][]{kollmeier2006}{}{}. By assuming a seed mass of $10^2\,\msun$ ($10^5\,\msun$), we find that a total quasar lifetime of $\approx600\,\mathrm{Myr}$ ($\approx350\,\mathrm{Myr}$) is required to grow the black holes under the assumption of Eddington-limited accretion. This is in good agreement with the estimate for $t_\mathrm{Q}$ obtained above\footnote{This estimate assumes that black holes grow at the Eddington limit for their entire history. The demographic properties of quasars at the present time, however, do not constrain black hole growth on a timescale larger than the inferred value of $t_\mathrm{Q}$. We can provide an alternative argument to link the quasar duty cycle to the growth of black hole mass by considering the characteristic luminosity of our quasar sample $L\gtrsim L_\mathrm{thr}$, and convert that to an accreted black hole mass by assuming a radiative efficiency of $\approx10\%$ and a total lifetime $t_\mathrm{Q}$. We get $\log_{10} M_\mathrm{BH}/\msun\approx8.7-9.7$ for $t_\mathrm{Q}=0.1-1\,\mathrm{Gyr}$, which again points to the fact that black holes can grow out to very high masses based on our inferred duty cycle.}.
This simple argument shows how studying the demographic properties of quasars (such as their abundance and clustering) can place indirect constraints on the formation and evolution history of SMBHs.

Alternative estimates for the quasar lifetime can be obtained by a number of other methods \citep[for an overview, see][]{martini_2004}{}{}. Interestingly, results from studies of the quasar proximity effect at high $z$ \citep[][]{khrykin2016, khrykin2019}{}{} paint a rather different picture than the one suggested here, finding values of the quasar lifetime that are several orders of magnitudes smaller \citep[see also][]{davies2019,eilers2021}{}{}. \citet{khrykin2021} compiled a set of HeII proximity zone measurements for $z\approx3-4$ quasars, and inferred a log-normal quasar lifetime distribution with a mean of $t_\mathrm{Q}\approx0.2\,\mathrm{Myr}$ and a standard deviation of $\approx0.8\,\mathrm{dex}$.  It is important to note, however, that proximity zone measurements are sensitive only to a fraction of the past quasar lightcurve (up to $\approx30\,\mathrm{Myr}$ for HeII). Clustering measurements, on the other hand, provide integral constraints on the total lightcurve emitted by quasars over the entire history of the Universe. In other words, they are only sensitive to the zeroth moment of the 
quasar lightcurve distribution (i.e., the aggregate probability of the lightcurve). The discrepancy between lifetime estimates for proximity zone sizes and clustering measurements, then, may suggest that quasar lightcurves exhibit non-trivial variations on timescales close to the ones probed by proximity zones. With this respect, exploring the full probability distribution associated with quasar lightcurves in the context of our quasar demographic model would provide a way to connect these very different observational probes of quasar activity in a single consistent picture. We will investigate this point in future work.

\subsection{Comparison with previous work}

The model presented in this work builds on a long-standing tradition of interpreting quasar observables via population modeling, i.e., by linking quasars to the well-known population of halos (or sometimes galaxies) according to some empirical/phenomenological prescriptions. Explaining the observed relative abundance of quasars at different luminosities (i.e., the QLF) within such frameworks has been achieved many times, with a large variety of empirical models and physical prescriptions 
employed \citep[e.g.,][]{efstathiou_rees1998,wyithe_loeb_2003,croton2009, conroy_white2013, fanidakis2013, veale2014, caplar2015, weigel2017,ren_trenti2021, trinity}. 
The bottom line is that the QLF is pretty straightforward to model starting from the hierarchical growth of structures predicted in the $\Lambda$CDM framework. On the other hand, the QLF alone does not place tight constraints on key properties
of quasars such as their black hole mass, accretion rate, lifetime, and host halo mass, not even in the context of redshift-dependent models \citep[e.g.,][]{wyithe2006, wyithe_loeb2009, veale2014}{}{}. Indeed, our analysis in Sec. \ref{sec:results_z4} (Fig. \ref{fig:results_mcmc}) suggests that a very wide variety of model parameters can be in good agreement with the QLF. As shown by \citet{veale2014}, alternative parametrizations would fare nearly equally well at all redshifts. The large uncertainties on the actual shape and normalization of the QLF that are due to the significant systematics involved in these measurements \citep[][]{kulkarni2019} exacerbate this issue, especially at high redshift. 

For this reason, considering the independent constraints coming from quasar clustering is extremely useful, as they provide constraints on the masses of the halos that are capable of hosting quasars. Reproducing the clustering of low-redshift ($z\lesssim2.5$) quasars has been shown to be possible both in empirical models \citep[e.g.,][]{kauffmann_haehnelt_2002,hopkins2007,croton2009, shankar2010_lowz,white2012,conroy_white2013, aversa2015, shankar2020}{}{}, semi-analytic models \citep[e.g.,][]{bonoli2007,fanidakis2013, oogi2016}{}{} and cosmological hydrodynamical simulations \citep[e.g.,][]{degraf2017}{}{}. All of these studies, however, show a significant tension with the clustering measurements at redshift $z\gtrsim3$.

The implications for the strong clustering measured by \citeS07 at $z\approx3-4$, in particular, have been discussed by \citealt{white_2008}, \citealt{wyithe_loeb2009}, and \citealt{shankar_2010}. \citet{white_2008} assume that the quasar luminosity-halo mass relation is linear, and find that the total number density of bright quasars can be reconciled with the linear bias measured by \citeS07 only if the scatter about this linear relation is very small ($\sigma\lesssim0.3~\mathrm{dex}$). They also find that their conclusions are strongly dependent on the specific functional form assumed for the linear bias-halo mass relation $b(M)$: the Mo \& White bias \citep[][]{mo_white1996, jing98}{}{} is found to be marginally compatible with data, whereas the Sheth, Mo, \& Tormen one \citep[][]{sheth_mo_tormen_2001}{}{} is inconsistent with the measured bias at the $\approx2\sigma$ level. 
Adopting the Sheth, Mo, \& Tormen functional form of $b(M)$ after showing that it is a better fit to N-body simulations, \citet{shankar_2010} interpret the \citeS07 data in the context of an evolutionary model for supermassive black holes, and they strengthen the conclusion that there is tension between the measured bias and the theoretical predictions at $z=4$. Similar results are found by \citet{wyithe_loeb2009}, who advocate for a contribution of a merger-driven bias to the $z=4$ clustering (see also \citealt{bonoli2010}; \citealt{cen2015} for a discussion of the impact of assembly bias on quasar clustering). 

Our work shares some similarities with the three studies mentioned above: we also assume a direct relation between quasar luminosity and halo mass and use the quasar clustering data to infer the specific shape of this relation.
Key conclusions of our analysis can also be found in these former attempts to explain the \citeS07 observations: in Sec. \ref{sec:results_lowz}, we find that bright quasars need to be hosted by very massive ($\log_{10} M /\msun \gtrsim 13$) halos, and, as a consequence, the quasar duty cycle is a significant fraction of unity ($\varepsilon_{\rm DC}\approx10-100\%$). 
In agreement with \citet{white_2008} and \citet{shankar_2010}, we conclude that a relatively small scatter ($\sigma\lesssim0.3$; Table \ref{tab:mcmc}) 
in the quasar luminosity-halo mass relation is necessary to explain the \citeS07 measurement. As also done by \citet{wyithe_loeb2009}, we adopt a more flexible parametrization of this relation by assuming that it can be non-linear, and find that a steep 
slope ($\gamma\gtrsim2$) achieves a much better fit to the data. 

The major novelty that our work brings to the understanding of this problem, however, does not reside in the interpretation of the results, but rather in the framework we use to build our model. 
As explained in Sec. \ref{sec:methods}, we extract the correlation function and the relative abundance of quasars directly from extremely large-volume cosmological N-body simulations, using a novel method to quickly compute the quasar auto-correlation function for any quasar-host mass distribution.
In this way, we can directly compare our predictions for the quasar projected correlation function with the \citeS07 observational data. Our model -- being based on N-body simulations -- naturally accounts for the non-linear contributions to quasar clustering that are essential to interpret the \citeS07 clustering measurements correctly, especially at scales $r\lesssim10-15\,\cMpc$. Using this approach, we thus achieve a much more solid data-model comparison, as we do not have to resort to the notion of large-scale linear bias, which is significantly uncertain for strongly biased systems \citep[][]{colossus_diemer2018}{}{}
and which discards the information about the shape and the physical features that lie in the \citeS07 data points. 

Indeed, according to the statistical analysis performed in Sec. \ref{sec:results_z4}, we find our model can match the data with a satisfactory level of accuracy (i.e., reduced chi-squared $\approx1$), suggesting that, despite being rather extreme, the \citeS07 data can be explained in the context of the standard framework in which clustering of dark matter halos is solely dictated by their mass, without the need to invoke any contributions from assembly/merger bias. We note that we cannot exclude, of course, that such a contribution is present. If that is the case, it would imply that the mass function of $z=4$ bright-quasar hosts may be somewhat less skewed towards very large halo masses ($\log_{10} M/\msun \gtrsim 13$). We leave an assessment of the role that merger bias plays in cosmological simulations to future work.

Finally, we note that several measurements of the characteristic mass of quasar-hosting halos at $z=2-4$ are available in the literature. They employ quasar-quasar (\citeS07; \citeE15; \citealt{timlin2018}) and quasar-galaxy \citep[][]{trainor_steidel2012,ikeda2015,garcia-vergara2017,he2018, garcia-vergara2019}{}{} clustering, as well as gas kinematics in the circumgalactic medium (CGM) of quasar-hosting galaxies \citep[][]{fossati2021,de_beer2023}{}{}. While the host halo masses predicted by these studies vary significantly, in the present work we have decided to focus on the \citeS07 and \citeE15 measurements of SDSS/BOSS quasars only, because these quasar samples are entirely spectroscopic and thus free from any low-redshift contaminants. However, the same analysis described in this paper could also be performed by taking into account the other clustering measurements mentioned above.

\subsection{Caveats and final remarks} \label{sec:discussion_caveats}

As shown schematically in Fig. \ref{fig:method_summary}, the results presented in this work depend on two key ingredients: the choice of the CLF, and the extraction of the halo mass function and the halo (cross-)correlation functions from cosmological N-body simulations. In the following, we will discuss the strengths and weaknesses of our method by considering these components in turn. Let us start with the latter: there are multiple sources of uncertainty in the final estimates we obtain for the halo mass function and the halo cross-correlation functions. First, despite the fact that the box sizes of the simulations employed here are among the largest ever run 
\citep[][]{angulo2022}{}{}, halos are so rare at the very massive end ($4-7\sigma$ peaks in the density field) that the results of simulations at these masses suffer from significant noise. In order to circumvent this issue -- and to extrapolate the results of simulations to the highest mass possible -- we used analytical functions to fit the data extracted from the simulations (Sec. \ref{sec:methods_sim}). These analytical fits, however, are not perfectly accurate and contribute some systematic errors to our final model predictions. 

Nonetheless, we believe that these sources of error can be neglected in our data-model comparison (Sec. \ref{sec:results}). This is because -- as also noted in Sec. \ref{sec:methods_sim} -- the observables we are trying to reproduce, i.e., the QLF and the projected correlation function, suffer from significant statistical and systematic uncertainties (as high as $\approx 100\%$ at $z=4$ and $\approx 30\%$ at $z=2.5$). In Sec. \ref{sec:simulations_hmf}-\ref{sec:simulations_corr} and Appendix \ref{sec:appendix_fitting}, we assess how well our fitting functions reproduce simulations, and show that their relative accuracy 
is generally $\lesssim5-10\%$ for both the halo mass function and the cross-correlation functions. For very small ($r\lesssim5\,\cMpc$) and very large ($r\gtrsim100\,\cMpc$) scales, measuring the cross-correlation functions in simulations is particularly challenging, especially at the high mass end. The small-scale behavior is highly affected by halo exclusion effects, whereas at large scales the finite size of the simulated boxes reduces the number of pairs, 
and the baryon acoustic oscillation (BAO)
peak makes the shape of correlation functions difficult to capture with our coarse radial bins. As a consequence, our fitting functions are also subject to larger errors at both of these scales. However, these errors do not have a significant impact on our final results, as observational data are also very uncertain at the same scales; for this very same reason, we have excluded the \citeS07 measurements at very large scales ($r>100\,\cMpc$) from our 
analysis (see Sec. \ref{sec:obs_data}).

As for the extrapolation of cross-correlations functions to masses higher than the ones that we can probe with our simulations, we have argued that such extrapolation is well motivated by considering the case of $z=2.5$ in Appendix \ref{sec:appendix_fitting_z2.5}. Furthermore, we note that the accuracy of this extrapolation does not have a significant impact on our results: this can be determined by looking at the QHMFs in Figure \ref{fig:discussion_redshift} (top panel). At both redshifts, the quasars hosted by halos whose mass is not well represented in simulations are only a small fraction of the total number of quasars (e.g., $\lesssim5\%$ at $z=4$). This implies that their actual contribution to the quasar auto-correlation function is negligible compared to the uncertainty in the data.

Other possible sources of uncertainty in our model predictions that we have not discussed yet are the cosmology assumed in the simulations and the exclusion of sub-halos in the creation of the halo cross-correlation functions. Cosmological parameters such as $\sigma_8$ and $\Omega_\mathrm{m}$ are predicted to have a significant effect on the collapse of structures in the standard $\Lambda$CDM model, and consequently on the spatial distribution of very massive halos at all reshifts. Studying the impact of these parameters on our final predictions for the clustering of quasars is beyond the scope of this work. Given the current large relative uncertainty on the data, however, we believe that including variations of the cosmological parameters in our inference procedure would have little effect on our final results. 

The exclusion of sub-halos is motivated by the fact that we are only considering clustering measurements at medium-large scales, which are not affected by the distribution of quasars inside a single halo 
-- the so-called one-halo term in HOD models \citep[e.g.,][]{cooray_sheth2001}{}{}. In principle, including the contribution of sub-haloes may boost the large-scale clustering too, as having multiple quasars living in the same dark matter halo implies a larger number of large-scale pairs. In practice, however, sub-haloes are less massive than centrals, and thus they do not tend to host very bright quasars according to the CLF found in Sec. \ref{sec:results}. We have included sub-haloes in some test runs and verified that large-scale clustering changes only at the percent level, and significant differences are only present at $r\lesssim1-2\,\cMpc$ even for the most massive halos. 
On top of that, we note that all of the effects discussed here go in the direction of an enhancement of the predicted clustering, and do not affect the main conclusion of this paper, i.e., that the very strong clustering measured at $z=4$ can be reproduced with standard assumptions of bright quasars inhabiting massive halos.

In this work, we have assumed one, very simple parametrization for the CLF. We believe that this simple framework is a strength of our model, as it provides very clear physical insight into the formation of quasars and their connection with the hierarchical growth of structures in the context of a $\Lambda$CDM universe. 
On the other hand, we have tested this basic parametrization on a relatively small amount of (very uncertain) data. We have done this on purpose: the main focus of this paper is on reproducing quasar clustering at $z=4$, and given the quality of the data we have at the present moment, a more sophisticated choice for the CLF would likely have been too flexible to be constrained. Indeed, we experimented with variations of the functional form assumed for the CLF, finding negligible improvements in terms of model-data comparison, and retaining the same fundamental conclusions in terms of quasar-hosting halo masses and duty cycles. Nonetheless, it is possible that extending our model to a larger/higher signal-to-noise ratio (S/N) dataset, e.g. at $z=0-2$, would be feasible only with more sophisticated parametrizations for the CLF. These may include, for example, a break in the quasar luminosity-halo mass relation, or a dependence of the active fraction, $f_\mathrm{on}$, and/or the scatter, $\sigma$, on halo mass. Some of these choices are particularly relevant, as they may be physically motivated, e.g., by a merger-driven accretion scenario for black hole activity. While similar changes in our CLF model are likely to have a direct impact on the inferred quasar luminosity-halo mass relation, we believe that the fundamental conclusions of our work regarding the evolution with redshift of (a) the characteristic mass of quasar-hosting halos and (b) the quasar duty cycle would remain substantially unchanged, because they are ultimately driven by the observed abundance of quasars and by their spatial distribution, and depend very little on our modeling details.
We leave a thorough examination of different prescriptions for the CLF for future work. In particular, we plan to apply our model to the multiple measurements of quasar clustering available at low-$z$ \citep[e.g.,][]{porciani_2004,croom2005, ross2009}{}{} as well as to the analyses of the dependence of clustering on luminosity at the same redshifts \citep[e.g.,][]{porciani_norberg2006,shen2009, eftekharzadeh2015}{}{}.

\section{Summary} \label{sec:conclusions}

We have introduced a novel framework that makes use of multiple cosmological N-body simulations to efficiently predict quasar observables such as the quasar luminosity function (QLF) and the quasar auto-correlation function. The halo mass function and the cross-correlation functions of halos with different masses are extracted from the dark-matter-only (DMO) versions of the FLAMINGO simulations and used to inform analytical fitting functions. 
These form the backbone of the model, which is then completed by the choice of a conditional luminosity function (CLF) that links halo masses to quasar luminosities. With these ingredients, we are able to predict the clustering and 
the luminosity function of quasars, as well as other key properties such as the mass distribution of quasar-hosting halos and the quasar duty cycle (Figure \ref{fig:method_summary}). 

We focus our analysis on the extremely strong clustering measured by \citet{shen2007} at $z\approx4$, with the goal of determining whether we can reproduce this measurement in the context of our model.
We use a simple parametrization for the CLF, assuming a power-law dependence of quasar luminosity on halo mass ($L\propto M^\gamma$) with a log-normal scatter $\sigma$. 
We fit the $z=4$ QLF and projected correlation function both independently and jointly, 
in order to gain insight into the best-fitting parameters for each of the cases considered. We then turn our attention to lower-$z$ data, and show that our model can also match the measurements of the same quantities at $z\approx2.5$ \citep[][]{ross2009,eftekharzadeh2015}{}{}, albeit with significantly different values of the model parameters. 

We summarise here the main findings of the analysis described above: 
\begin{itemize}
    \item Quasar clustering and abundance measurements at $z\approx4$ require quasars to reside in the most massive halos at that redshift, with a characteristic mass of $\log_{10} M/\msun\gtrsim 13$ (Figure \ref{fig:results_overview}). This implies that the relation between quasar luminosity and halo mass ($L-M$) is highly non-linear ($\gamma\gtrsim2$) with a very small amount of scatter ($\sigma\lesssim0.3~\mathrm{dex}$). 
    \item Many different combinations of model parameters can achieve a good fit to the measured QLF at $z\approx4$ (Figure \ref{fig:results_mcmc}). This is because very different empirical prescriptions for the quasar luminosity-halo mass relation (e.g., large scatter and shallow slope or vice-versa) are able to map the exponentially declining end of the halo mass function into the shallower bright end of the QLF. However, the only set of parameters which is also compatible with clustering measurements is the one mentioned above (i.e, a highly non-linear $L-M$ relation with very small scatter), as an increase in the scatter would lower the characteristic mass of quasar-hosting halos, and thus decrease the clustering predicted by our model.  
    \item In order to match the total number density of bright $z\approx4$ quasars in models in which quasars reside in sufficiently high halo masses to reproduce the observed clustering, the active fraction of quasars ($f_{\rm on}$) has to be close to unity. This implies that high-$z$ quasars shine for a large fraction of the Hubble time, with a duty cycle in the range $\varepsilon_{\rm DC}=10-60\%$. In turn, this duty cycle results in a large total quasar lifetime $t_\mathrm{Q}\approx10^8-10^9~\mathrm{yr}$, consistent with the standard picture of black hole growth in the young universe. 
    \item The steep $z\approx4$ relation between quasar luminosity and halo mass contrasts with the well-known prediction of a flattening in the stellar mass-halo mass relation at high mass at every epoch \citep[e.g.][]{behroozi2013}{}{}. 
    This implies that in very massive high-$z$ halos -- while the star formation may have been quenched already -- the supermassive black hole at the center of the galaxy needs to be either over-massive and/or highly accreting. This may have an impact on the shape and normalization of the black hole mass-galaxy mass relation at high redshift (see e.g., \citealt{maiolino2023, stone2023}). 
    \item Furthermore, the extremely small scatter ($\sigma\approx0.1-0.3~{\rm dex}$) inferred for the $L-M$ relation at $z\approx4$ points to some physical processes enforcing a tight relationship between quasars and their dark matter halo hosts. In other words, the relation between black hole mass and stellar and/or halo mass, together with the distribution of Eddington ratios, all conspire to yield a remarkably low scatter. 
    \item The clustering and relative abundance of quasars at lower redshift ($z\approx2.5$) can be explained by the same parametric relation between quasar luminosity and halo mass. However, the parameters describing this relation show a significant evolution with redshift (Figure \ref{fig:results_mcmc_redshift}): the slope of the $L-M$ 
    is significantly shallower ($\gamma\approx1.15$) than at $z\approx4$, and the scatter larger ($\sigma\approx0.5~\mathrm{dex}$).
    \item Overall, our comparison between $z\approx2.5$ and $z\approx4$ reveals two radically different pictures in terms of the connection between quasars and their host halo population (Figure \ref{fig:discussion_redshift}). High-$z$ ($z\approx4$) quasars are hosted by very massive halos, with a very large occupation fraction (i.e., a large fraction of these halos host bright quasars at any given time). At lower redshift ($z\approx2.5$), instead, quasars reside in halos with a broad range of masses, with the bulk of the population being characterized by relatively common, $\log_{10} M /\msun \approx 12.5$ mass halos. As a consequence, only a small fraction of low-$z$ quasars are actively shining at any given moment, with a quasar duty cycle of $\varepsilon_{\rm DC}\approx0.5\%$. These conclusions are consistent with the standard picture of ``cosmic downsizing'' of quasars and AGN \citep[e.g.,][]{merloni2004, scannapieco2005, fanidakis2012}{}{}, as the bulk of the quasar population is hosted by progressively smaller halos as redshift decreases.
\end{itemize}

The framework presented here can be readily applied to interpret quasar clustering measurements at all redshifts. In particular, focusing on very high redshift is especially interesting in light of the fact that the large-scale environment of very bright quasars has been proven hard to pinpoint in the early universe \citep[e.g.,][]{fan2022}{}{}. 
For example, \citet{arita2023} recently measured the quasar auto-correlation function at $z\approx 6$ for the first time, finding results broadly consistent with the very strong clustering measured at $z\approx4$. On top of that, several JWST programs such as ASPIRE \citep[][]{aspire_wang2023}{}{} and EIGER \citep{eiger_kashino2023,eiger_eilers2023} are starting to deliver 
measurements of quasar clustering by probing the distribution of line emitters around bright, $z\approx6-7$ quasars. Connecting the framework presented here to the upcoming quasar-galaxy cross-correlation measurements from JWST will offer a clear and comprehensive picture of the large-scale environments in which the first quasars formed (see also \citealt{costa2023} for an alternative approach). 

As suggested by the results obtained in this work, interpreting quasar properties within a consistent framework that takes into account both their demographics and their spatial distribution can give great insight into the relationship between the hierarchical growth of structures in the universe and the evolution of supermassive black holes over cosmic time.

\section*{Acknowledgements}

We are grateful to the FLAMINGO team for making their dark matter only simulations available. 
We acknowledge helpful conversations with the ENIGMA group at
UC Santa Barbara and Leiden University. 
EP is grateful to Roi Kugel for helpful discussion, and to Molly Wolfson, Shane Bechtel, and Silvia Onorato for comments on an early draft of this paper. 
JFH and EP acknowledge support from the European Research Council (ERC) under the European
Union’s Horizon 2020 research and innovation program (grant agreement No 885301).
This work is partly supported by funding from the European Union’s Horizon 2020 research and innovation programme under the Marie Skłodowska-Curie grant agreement No 860744 (BiD4BESt). 
This work used the DiRAC@Durham facility managed by the Institute for 
Computational Cosmology on behalf of the STFC DiRAC HPC Facility
(\url{www.dirac.ac.uk}). The equipment was funded by BEIS capital funding via
STFC capital grants ST/K00042X/1, ST/P002293/1, ST/R002371/1 and ST/S002502/1,
Durham University and STFC operations grant ST/R000832/1. DiRAC is part of the
National e-Infrastructure.

\section*{Data Availability}
The derived data generated in this research will be shared on reasonable requests to the corresponding author.



\bibliographystyle{mnras}
\bibliography{biblio} 




\appendix

\section{Obtaining the quasar auto-correlation from the halo cross-correlation functions} \label{sec:cross-corr}

Let us consider a stochastic process $\mathcal{N}^{(q)}$, describing -- in our case of interest -- the spatial distribution of quasars. This distribution is discrete: following \citet{peebles_book}, we divide the volume of interest into infinitesimal elements $\delta V_i$, and -- given the average quasar density $\Bar{n}_q$ -- we can write the probability of having a quasar in the volume element $\delta V_1$ as:
\begin{equation}
    \delta P_1 = \langle \mathcal{N}^{(q)}_1 \rangle = \Bar{n}_q \,\delta V_1.
\end{equation}
Similarly, we define the two-point correlation function, $\xi(r_{12})\equiv\xi_{12}$, via the probability of having a quasar in the volume element $\delta V_1$ and another one in the volume element $\delta V_2$:
\begin{equation}
    \delta P_{12} = \langle \mathcal{N}^{(q)}_1\mathcal{N}^{(q)}_2 \rangle \equiv \Bar{n}_q^2 \,\delta V_1\delta V_2\left(1+\xi_{12}\right). \label{eq:xi12}
\end{equation}
We introduce now the continuous number density field $n^{(q)}(\mathbf{x})$, which we define via the expression:
\begin{equation}
    \delta P_{12} = \langle \mathcal{N}^{(q)}_1\mathcal{N}^{(q)}_2 \rangle \equiv \langle n^{(q)}(\mathbf{x}_1) n^{(q)}(\mathbf{x}_2)  \rangle \,\delta V_1\delta V_2,
\end{equation}
and write this equation in terms of the density contrast field $\delta^{(q)}$ -- defined as $n^{(q)}(\mathbf{x}) =  \Bar{n}_q\,\left(1+\delta^{(q)}(\mathbf{x})\right)$:
\begin{equation}
    \delta P_{12} = \Bar{n}_q^2 \,\delta V_1\delta V_2 \left(1+\langle\delta^{(q)}(\mathbf{x}_1)\delta^{(q)}(\mathbf{x}_2)\rangle\right).
\end{equation}
By comparing this to eq. \ref{eq:xi12}, we find that the correlation function can also be expressed as:
\begin{equation}
        \xi_{12} = \langle\delta^{(q)}(\mathbf{x}_1)\delta^{(q)}(\mathbf{x}_2)\rangle \equiv \langle\delta^{(q)}_1\delta^{(q)}_2\rangle .
\end{equation}
We want to split the different contributions of quasars to the density field $n^{(q)}$ based on the mass of their host halos. We introduce a set of continuous fields $\{n^{(h)}(M_k)\}$, which represent the distributions of halos for different mass bins centered on $M_k$ and with a width $\Delta M$. 

We now make the key hypothesis that the distribution of quasars with a host halo mass in the range $[M_k-\Delta M/2, M_k+\Delta M/2]$ is an unbiased tracer of the underlying distribution of halos, $n^{(h)}(M_k)$. In other words, the quasars at a given host halo mass are just undersampling the distribution of halos, and they are thus described by the same stochastic process. This is the case if the presence of a quasar depends solely on the mass of its host. Thus, we can write the quasar distribution, $n^{(q)}$ in terms of the distributions of halos with different masses by simply weighing them by the relative number of quasars at those masses:
\begin{equation}
    n^{(q)}=\sum_k p_k\,n^{(h)}(M_k),
\end{equation}
where $p_k$ represents the probability that a quasar has a host-halo mass in the bin $M_k$.
Using the ``quasar-host mass function'' (QHMF) introduced in Sec. \ref{sec:methods_basic} (eq. \ref{eq:qhmf}), we can express this probability as ($\Bar{n}_{q,k}$ is the average number density of quasars in the host mass bin $M_k$):
\begin{equation}
     p_k = \frac{\Bar{n}_{q,k}}{\Bar{n}_q} = \frac{n_{\rm QHMF}(M_k)\,\Delta M}{\int_0^\infty n_{\rm QHMF}(M)\,\d M}.
\end{equation}
Introducing the overdensity definitions for the distributions at different masses, $n^{(h)}(M_k) = \Bar{n}_{q,k}\,\left(1+\delta^{(h)}(M_k)\right)$
we can write:
\begin{equation}
\begin{split}
    \langle n^{(q)}_1 n^{(q)}_2\rangle &=  \sum_j\sum_k p_jp_k \Bar{n}_q^2\, \left(1+\langle\delta_1^{(h)}(M_j)\delta_2^{(h)}(M_k)\rangle\right)=\\
    &= \Bar{n}_q^2 \left(1+\sum_j\sum_k p_jp_k \langle\delta_1^{(h)}(M_j)\delta_2^{(h)}(M_k)\rangle\right),
\end{split}
\end{equation}
where we have made use of the fact that $\sum_kp_k=1$. Introducing the cross-correlation functions for halos of different masses, $\xi^{(h)}_{12}(M_j,M_k)=\langle\delta_1^{(h)}(M_j)\delta_2^{(h)}(M_k)\rangle$, we can express the quasar auto-correlation as:
\begin{equation}
    \xi_{12} = \langle \delta^{(q)}_1\delta^{(q)}_2 \rangle =  \frac{\langle n^{(q)}_1 n^{(q)}_2 \rangle}{\Bar{n}_q^2}-1 = \sum_{j,k} p_jp_k\, \xi^{(h)}_{12}(M_j,M_k).
\end{equation}
This proves eq. \ref{eq:quasar_corr_func}, which relates the quasar auto-correlation function $\xi(r_{12})\equiv\xi_{12}$ to the cross-correlation functions of halos with different masses, $\xi_h(M_j, M_k;r_{12})\equiv \xi^{(h)}_{12}(M_j,M_k)$.

\section{Fitting the cross-correlation terms from simulations} \label{sec:appendix_fitting}

In this Section, we provide details on the fitting we perform to the values of the halo cross-correlation functions, $\xi_h(M_j, M_k; r)$, extracted from simulations. As described in Sec. \ref{sec:methods_sim} and \ref{sec:simulations_corr}, we extract these values from two simulations with box sizes equal to $L=2800\,\cMpc$ and $L=5600\,\cMpc$, respectively. For each simulation, we consider only halos in a specific range of masses, so that all the mass bins considered are populated by a sufficient number of well-resolved halos. In particular, we set the following ranges (Sec. \ref{sec:methods_sim}): $\log_{10} M/\msun = 11.5-13.0$ for $L=2800\,\cMpc$, and $\log_{10} M/\msun = 12.5-13.5$ for $L=5600\,\cMpc$. We choose a bin width of $0.25$ in $\log_{10} M$, so that we have $6\times 6$ cross-correlation terms for $L=2800\,\cMpc$, and $4\times 4$ cross-correlation terms for $L=5600\,\cMpc$. Note that the masses $M_j$ and $M_k$ in the expression $\xi_h(M_j, M_k; r)$ do not refer to the center of their respective bins, but rather to the median value of the halo mass function in those bins. 

Our goal is then to find a single analytical description of these cross-correlation functions that can represent the two simulations simultaneously. In order to do that, we first divide all the cross-correlation functions $\xi_h(M_j, M_k; r)$ by a reference correlation $\xi_\mathrm{ref}(r)$; in formulae, we define $\rho(M_j, M_k; r)$ to be:
\begin{equation}
    \rho(M_j, M_k; r) = \xi_h(M_j, M_k; r) / \xi_\mathrm{ref}(r).
\end{equation}
In this way, we hope that $\rho(M_j, M_k; r)$ will be only marginally dependent on the scale $r$. We set $\xi_\mathrm{ref}(r)\equiv\xi_h(\Tilde{M}, \Tilde{M}; r)$, with $\Tilde{M}$ representing the $\log_{10} M = 12.5-12.75$ bin. This choice is arbitrary, but it is made to ensure that the mass bin sits in the overlap between the mass ranges of the two different simulations we use.
We also attempt to minimize any dependences of the cross-correlation functions on cosmology and redshift by expressing all the masses in terms of peak heights $\nu(M)$ (see also Sec. \ref{sec:simulations_hmf}).

Finally, we fit a 3-d polynomial $\rho_\mathrm{fit}(\nu_j, \nu_k, r)$ to the values extracted from the simulations. We empirically find that a second-degree polynomial in mass and third-degree in the radial dimension fits the data points well enough and at
the same time attains a smooth behavior with respect to all three variables (i.e., we prevent overfitting). The errors on the data points are assigned based on the Poisson statistics of the pair counts (eq. \ref{eq:pair_counts}). As also done in Sec. \ref{sec:simulations_hmf}, we weigh the errors associated with the two simulations differently in order to achieve a better fit. In particular, we double the values of the Poisson errors for the $L=2800\,\cMpc$ simulation, and we halve the ones associated with the $L=5600\,\cMpc$ box.

\begin{figure*}
	\centering
	\includegraphics[width=\textwidth]{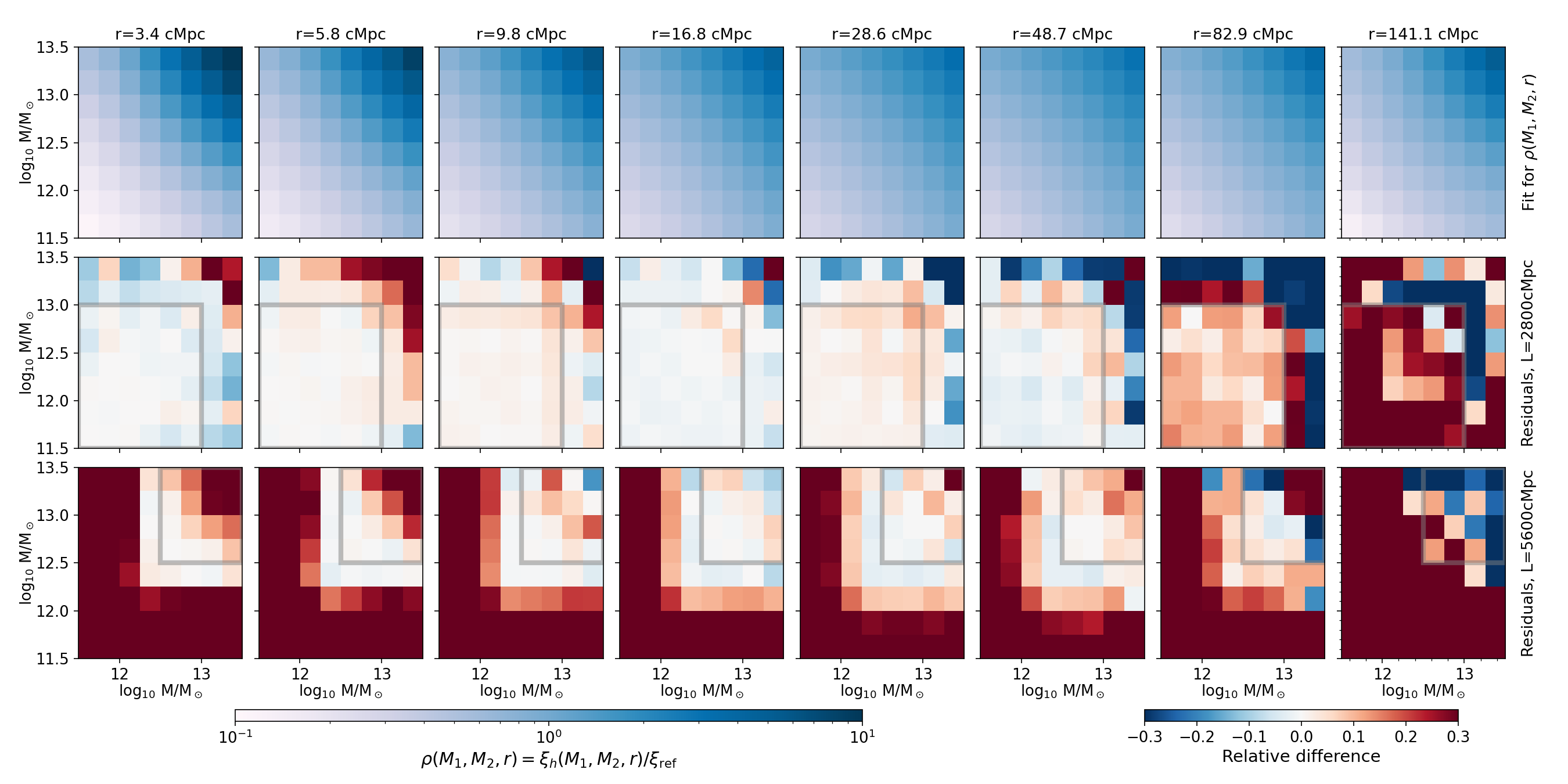}

	 \caption{ Results for the fitting of the cross-correlation terms $\rho(M_1, M_2, r)$ (see Appedix \ref{sec:appendix_fitting} for definitions). The top row shows the fitting function $\rho_\mathrm{fit}(M_1, M_2, r)$ as a function of the two masses $M_1$ and $M_2$ for different values of the distance $r$. The second and third rows show the relative difference between the fits and the values measured from the simulations ($L=2800\,\cMpc$ and $L=5600\,\cMpc$, respectively). Mass ranges that are adopted for the fitting in each simulation are highlighted by the gray boxes in each mass-mass plane.  
  \label{fig:corr_fitting_2d_z4}
 	}
\end{figure*}

\begin{figure*}
	\centering
	\includegraphics[width=\textwidth]{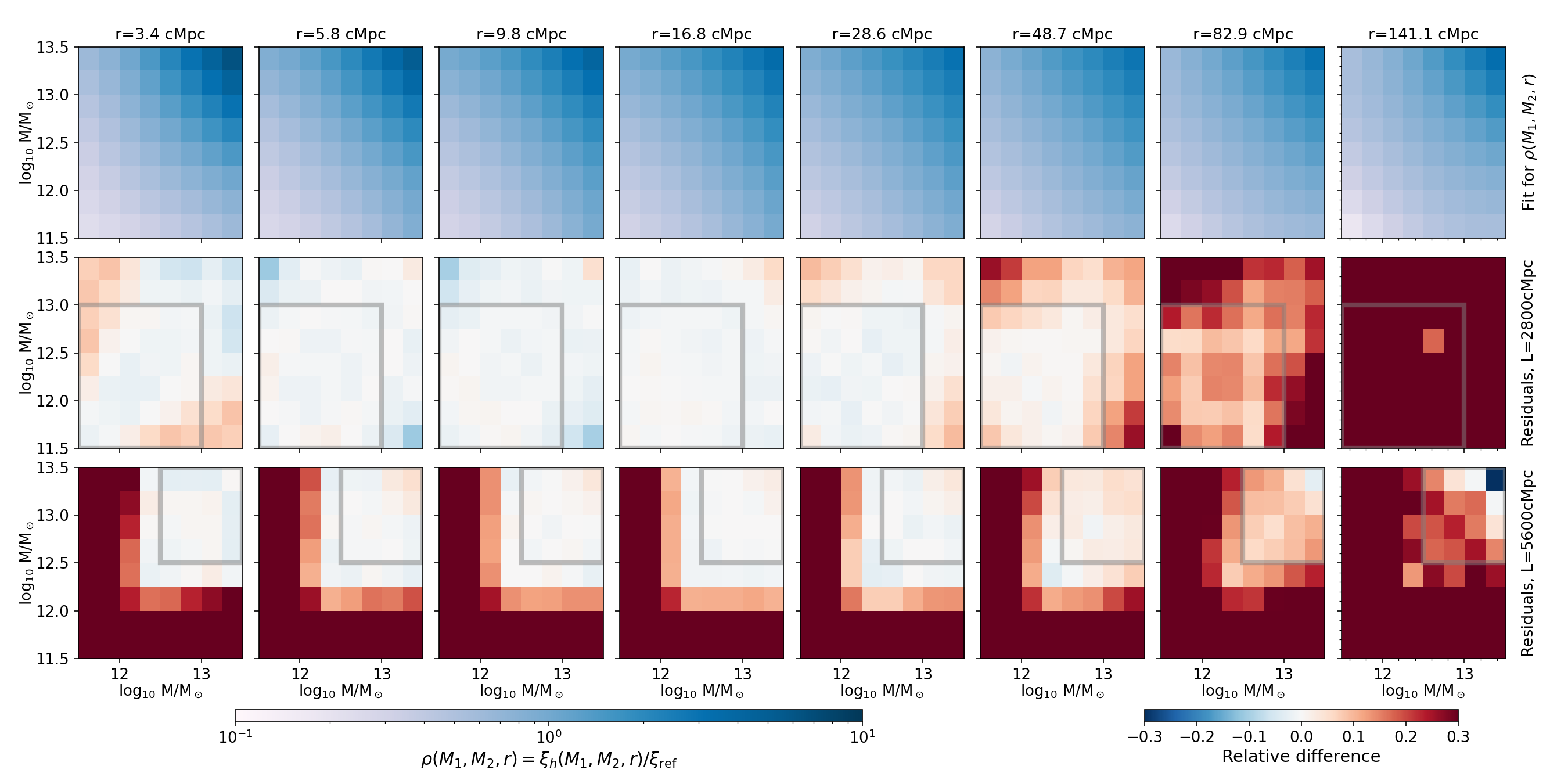}

	 \caption{Same as Figure \ref{fig:corr_fitting_2d_z4}, but for redshift $z=2.5$.
  \label{fig:corr_fitting_2d_z2}
 	}
\end{figure*}

Figures \ref{fig:corr_fitting_2d_z4}--\ref{fig:corr_fitting_2d_z2} show the results of the fitting for the two redshifts considered in this work: $z=4$ (Fig. \ref{fig:corr_fitting_2d_z4}) and $z=2.5$ (Fig. \ref{fig:corr_fitting_2d_z2}). The first row of each plot displays the resulting fitting function $\rho_\mathrm{fit}(\nu(M_j), \nu(M_k), r)$. Each panel in this row shows the values of $\rho_\mathrm{fit}(\nu(M_j), \nu(M_k), \Bar{r})$ as a function of the two masses $M_j$ and $M_k$ at a different scale $\Bar{r}$. The second and third rows show the relative differences ($\rho/\rho_\mathrm{fit} -1$) between our fit and the two simulations considered. The mass ranges that are selected in each simulation are shown as grey boxes in the 2-d mass planes. According to these figures, our simple analytical framework can describe the behavior of cross-correlation functions in a wide mass range with a good degree of accuracy ($\lesssim 5-10\%$). Notable exceptions to this can be found for very high masses ($\log_{10} M/\msun \gtrsim 13.2-13.3$) and very large scales ($r\gtrsim100\,\cMpc$). However, these are both expected, as both at large masses and large scales correlation functions are difficult to measure in simulations. Further discussion of this can be found in Sec. \ref{sec:simulations_corr} and Sec. \ref{sec:discussion_caveats}.
Similar conclusions on the quality of our fits can be drawn by looking at the right panels of Figure \ref{fig:simulations} and Figure \ref{fig:simulations_z2}, where predictions from our fitting functions are compared to auto-correlation functions extracted from the simulations.

\section{Halo mass function and correlation functions for redshift ${\small z}=2.5$} \label{sec:appendix_fitting_z2.5}

\begin{figure*}
	\centering
	\includegraphics[width=0.48\textwidth]{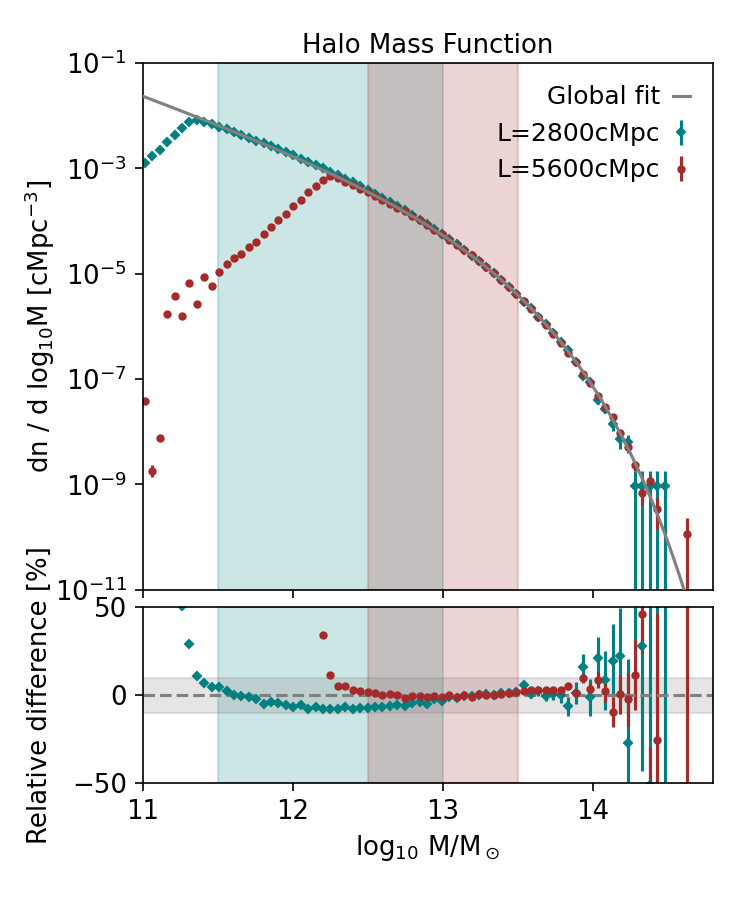}
 	\includegraphics[width=0.48\textwidth]{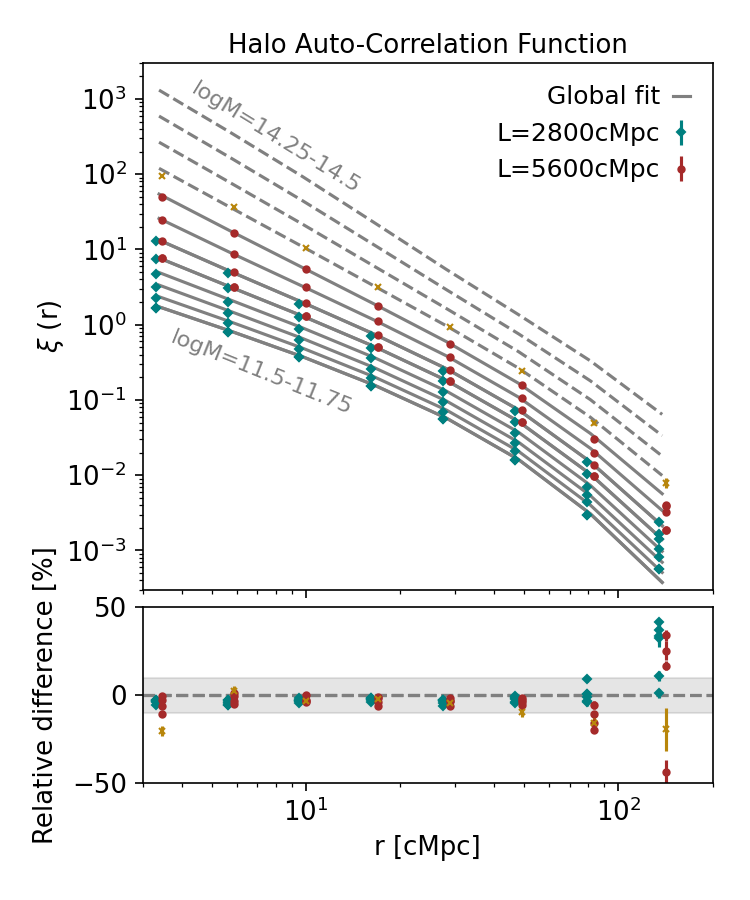}

	 \caption{Same as Fig. \ref{fig:simulations} but for the snapshots at redshift $z=2.5$. Golden crosses in the right panel represent the auto-correlation functions measured in the mass bin $\log_{10} M /\msun = 13.5-13.75$, in the $L=5600\,\cMpc$ simulation. This is used as a benchmark to assess how well our fits (dashed grey lines) can be extrapolated to higher masses. \label{fig:simulations_z2}
 	}
\end{figure*}

In the main text (Sec. \ref{sec:simulations_hmf}--\ref{sec:simulations_corr}), we discussed our predictions for the halo mass function and the halo (cross-)correlation functions at $z=4$. We show here the same results for the other redshift that we consider in our analysis, $z=2.5$. In Figure \ref{fig:simulations_z2} (left panel), we show the halo mass function extracted from the two simulations ($L=2800\,\cMpc$ and $L=5600\,\cMpc$ in teal and red, respectively), as well as our fitting function (eq. \ref{eq:hmf_fit}-\ref{eq:hmf}, gray line). The best-fitting parameter values for $z=2.5$ are: $A=0.464$; $a=3.43$; $b=0.847$; $c=1.31$. 
Note that for the fitting we employ the same mass ranges as we used for $z=4$ (Table \ref{tab:sim}). This choice is clearly sub-optimal, as halos are much more abundant at lower-$z$, and therefore mass bins with $\log_{10} M/\msun > 13.5$ are well populated even for the smallest box considered ($L=2800\,\cMpc$). However, we choose to not take into account masses larger than $\log_{10} M/\msun > 13.5$ for the fitting so that we can benchmark how well our fitting function fares if extrapolated to masses larger than this limit. In this way, we can test whether our fitting framework is valid to interpret the behavior of the halo mass function up to masses higher than the ones we can simulate. 
As shown in Fig. \ref{sec:appendix_fitting_z2.5} (left panel), the trend of the halo mass function at large masses is well described by the extrapolation of our fitting up to $\log_{10} M/\msun \gtrsim 14$; at higher masses, halos become very rare even at $z=2.5$, and the halo mass function becomes quite noisy and its behavior highly uncertain.  

The right panel of Figure \ref{fig:simulations_z2} shows the halo auto-correlation functions for different mass bins obtained both from simulations (colored points) and from our fitting function (gray lines; see Appendix \ref{sec:appendix_fitting}). We use 8 mass bins ranging from $\log_{10} M/\msun = 11.5$ to $\log_{10} M/\msun = 13.5$ and with a width of 0.25 dex. Lower mass bins correspond in Fig. \ref{sec:appendix_fitting_z2.5} to lower values of the correlation functions, and vice-versa.
Once again, the mass ranges employed for our fitting are the same for both $z=2.5$ and $z=4$, and do not go higher than $\log_{10} M/\msun =13.5$. This gives us the possibility of testing how the extrapolation of $\xi_{h, {\rm fit}}(M_j, M_k; r)$ fares at larger masses. 
As explained in Sec. \ref{sec:simulations_corr}, ensuring that our theoretical framework can be extended to very high masses ($\log_{10} M/\msun \approx14$) is quite important, as -- especially at $z=4$ -- a significant fraction of quasars are hosted by this population of massive halos that is not well represented in our simulations.  

Extrapolations from our fits are shown in Fig. \ref{fig:simulations_z2} (right panel) with dashed lines. We also extract halo auto-correlation functions from the $L=5600\,\cMpc$ box for the mass bin $\log_{10} M /\msun = 13.5-13.75$; we show these values with golden crosses in Fig. \ref{fig:simulations_z2} (right panel). We see that the extrapolation agrees with simulations at the same level as the points that are used for fitting, with the only exceptions being very small ($r\lesssim 5 \,\cMpc$) and very large ($r\gtrsim 100\,\cMpc$) scales. Further discussion on the implications of these results can be found in Sec. \ref{sec:discussion_caveats}.

\bsp	
\label{lastpage}
\end{document}